\documentclass[10pt,aps,twocolumn,prd,notitlepage,amssymb,amsmath,floatfix,nofootinbib,superscriptaddress]{revtex4-1}
\usepackage{epsfig}
\usepackage{bm}
\usepackage{amssymb}
\usepackage{amsmath}
\usepackage{color}
\usepackage{xcolor}
\usepackage[colorlinks,linkcolor=blue,anchorcolor=black,citecolor=blue]{hyperref}
\usepackage{multirow}

\allowdisplaybreaks[4]

\mathchardef\mhyphen="2D

\begin{document}

\title{Vector meson's spin alignments in high energy reactions}

\author{Jin-Hui Chen}
\email{chenjinhui@fudan.edu.cn}
\affiliation{Key Laboratory of Nuclear Physics and Ion-beam Application (MoE), Institute of Modern Physics, Fudan University, Shanghai 200433, China}
\author{Zuo-Tang Liang}
\email{liang@sdu.edu.cn}
\affiliation{Institute of Frontier and Interdisciplinary Science, Key Laboratory of Particle Physics and Particle Irradiation (MOE), Shandong University, Qingdao, Shandong 266237, China}
\author{Yu-Gang Ma}
\email{mayugang@fudan.edu.cn}
\affiliation{Key Laboratory of Nuclear Physics and Ion-beam Application (MoE), Institute of Modern Physics, Fudan University, Shanghai 200433, China}
\author{Xin-Li Sheng}
\email{sheng@fi.infn.it}
\affiliation{INFN-Firenze, 50019 Sesto Fiorentino FI, Italy}
\author{Qun Wang}
\email{qunwang@ustc.edu.cn}
\affiliation{Department of Modern Physics, University of Science and Technology of China, Hefei, Anhui 230026, China}
\affiliation{School of Mechanics and Physics, Anhui University of Science and Technology, Huainan, Anhui 232001, China}

\begin{abstract}
The global spin alignment of vector mesons has been observed by the STAR collaboration at the Relativistic Heavy Ion Collider (RHIC) at Brookhaven National Laboratory (BNL).
It provides a unique opportunity to probe the correlation between the polarized quark and antiquark in the strongly coupled quark-gluon plasma (sQGP) produced in relativistic heavy ion collisions, opening a new window to explore the properties of sQGP. 
In addition, spin alignments of vector mesons have also been observed in other high-energy particle collisions such as $e^+e^-$ annihilations at high energies where hadron production is dominated by quark fragmentation mechanism. The results obtained are quite different from those obtained in heavy ion collisions where quark coalescence/combination mechanism dominates suggesting strong dependence on hadronization mechanisms. So comprehensive studies in different hadronization processes are needed.
In this article, we present a brief review of theoretical and experimental advances in the study of vector meson's spin alignments 
in a variety of high-energy particle collisions, with emphasis on hadronization mechanisms.\\

\noindent
{\bf Keywords}: Vector meson, spin alignment, global polarization, relativistic heavy ion collisions, fragmentation\\
\noindent
{\bf PACS}: 12.38.Mh,12.38.–t, 13.88.+e, 13.66.Bc, 25.75.–q, 25.75.Nq

\end{abstract}

\maketitle

\section{Introduction}

Quantum Chromodynamics (QCD) is the fundamental theory for strong interaction that is responsible for binding quarks and gluons together to form protons, neutrons, and other hadrons. Relativistic Heavy Ion Collisions (RHIC) are a powerful tool for studying QCD and properties of strongly interacting matter under extreme conditions \cite{Braun-Munzinger:2015hba,Florkowski:2018fap,Ko,Rapp,Zhang1,ZhaoJ}. In these collisions, heavy nuclei are accelerated to nearly light speed and then collide to achieve immense energy density that recreates conditions similar to those just microseconds after the Big Bang, allowing scientists to study quark-gluon plasma (QGP) - a state of matter in which quarks and gluons are decoupled from hadrons~\cite{Pandav:2022xxx,HeWB,Chen:2023mel,MaYG,PangLG,Sun1,Chen1}.

Spin is an intrinsic form of angular momentum carried by elementary particles, and it is a (pseudo-)vector quantity that can point to different directions. As elementary particles of strong interaction, quarks and gluons carry spins. As composite particles of strong interaction, many hadrons also carry spins. The spin polarization refers to the alignment of the spin along a specific direction (called spin quantization direction). Spin degrees of freedom have been playing an important role in the development of modern physics since its discovery in 1925. 
The global polarization effect (GPE) of the quark gluon plasma (QGP) produced in high energy heavy-ion collisions is a new spin effect in particle and nuclear physics.
The theoretical prediction~\cite{Liang:2004ph,Liang:2004xn,Gao:2007bc} was made almost two decades ago and attracted immediate attention~\cite{Voloshin:2004ha,Betz:2007kg,Ipp:2007ng,Becattini:2007sr,Liang:2007ma,Chen:2007zzq,Selyuzhenkov:2007ab,STAR:2008lcm,STAR:2007ccu}.
However the enthusiasm was soon dampened by STAR's earlier attempts~\cite{Chen:2007zzq,Selyuzhenkov:2007ab,STAR:2008lcm,STAR:2007ccu}
that gave null results for global $\Lambda$-hyperon polarization as well as global spin alignments of vector mesons due to limited statistics in data at that time. 
The earlier results show that the GPE, even if it exists, could be very small, so it is unclear if it is 
within the scope and resolution of current experiments. 
The enthusiasm was aroused by non-vanishing results for the GPE for $\Lambda$ hyperons in STAR's beam-energy-scan experiments~\cite{STAR:2017ckg}, which showed that the GPE decreases monotonically with collision energies.
Now the GPE has grown to be a sub-field in heavy-ion collisions, see Refs.~\cite{Wang:2017jpl,Liang:2019clf,Gao:2020lxh,Huang:2020dtn,Gao:2020vbh,Becattini:2020ngo,Becattini:2022zvf,Becattini:2024uha} for recent reviews.

Another surprise came five years later from STAR's measurements on the global spin alignment of $\phi$ mesons with high statistics~\cite{STAR:2022fan}. 
The STAR's results~\cite{STAR:2022fan} seem to conflict with hyperon's polarization as predicted in Ref.~\cite{Liang:2004xn} that the vector meson's spin alignment is proportional 
to quark polarization squared and thus should be much smaller than that observed in STAR's experiment~\cite{STAR:2022fan}. 
So STAR's results are definitely non-trivial and may have deep implication for hadronization mechanism and properties of QCD~\cite{Yang:2017sdk,Sheng:2019kmk,Sheng:2022ffb,Sheng:2022wsy}.
When a system of particles exhibits spin polarization, it means that the spins of these particles are aligned more often in a particular direction than randomly oriented.
Now the global spin alignment of vector mesons has attracted a broad interest in experimental and theoretical communities~\cite{Wang:2023fvy,Chen:2023hnb,Xin-Li:2023gwh,Jian-Hua:2023cna,Li-Juan:2023bws,Chen:2024bik}.
Spin polarization in the context of QCD is a critical aspect of understanding the internal structure and dynamics of hadrons. Through both theoretical and experimental approaches, physicists aim to unravel the complexities of how quarks and gluons contribute to the spin and other properties of hadrons.
Several theoretical interpretations have been proposed and new measurements are underway.

In addition to relativistic heavy-ion collisions, the vector meson's spin alignment has also been studied in other high energy processes such as $e^+e^-$, $e^-p$ and $pp$ collisions~\cite{DELPHI:1997ruo,OPAL:1997vmw,OPAL:1997nwj,OPAL:1999hxs,ZEUS:1999wqh,NOMAD:2006kuc,STAR:2008lcm}. 
The results show quite different features and theoretical efforts have also been made to describe them~\cite{Anselmino:1997ui,Anselmino:1998jv,Anselmino:1999cg,Xu:2001hz,Xu:2003fq,Chen:2016moq,Chen:2016iey,Chen:2020pty}. 
It is therefore desirable to summarize these experimental and theoretical results 
and make comparison among different high energy reactions as guidance for future studies.

Firstly, we briefly review spin density matrices for vector mesons as well as measurement methods 
for the spin alignment, and then review experimental results in different high energy collisions. 
Secondly, we summarize theoretical approaches in two different hadronization mechanisms as well as 
the linear response theory for the spin alignment of vector mesons in thermalized QGP. 
Finally, we present a short summary and outlook.

\section {Spin density matrices and measurement methods for spin alignment}

\subsection{Spin density matrices} 
The spin polarization of particles produced in high energy reaction can be described by the spin density matrix $\hat\rho$. For particles with spin-$1/2$ such as quarks and anti-quarks, $\hat\rho_q$ is a $2\times 2$ Hermitian matrix which can be expanded as
\begin{align}
\hat\rho _q=\frac{1}{2}(1+\vec P_q\cdot\vec \sigma), 
\label{eq:rho1/2}
\end{align}
where $\vec\sigma =(\sigma_x,\sigma_y,\sigma_z)$ are Pauli matrices, and $\vec P_q={\rm Tr}(\vec\sigma\hat\rho_q)$ is the mean spin polarization vector in the quark's rest frame.

For spin-1 particles, the spin density matrix is a $3\times 3$ Hermitian matrix.
With the spin quantization axis specified, the spin states are denoted as $|jm\rangle$ and 
the spin density matrix can be put into the form~\cite{Bacchetta:2000jk} 
\begin{align}
\hat\rho _V=&\left(
\begin{array}{ccc}
\rho_{11} & \rho_{10}& \rho_{1-1}  \\
\rho_{01} & \rho_{00}& \rho_{0-1}  \\
\rho_{-11} & \rho_{-10}& \rho_{-1-1}
\end{array} \right)\nonumber\\
=&\frac{1}{3}+\frac{1}{2}P_{i}\Sigma_{i}+T_{ij}\Sigma_{ij} ,
\label{mvsd-decomp}
\end{align}
where $i,j$=1,2,3, and $\Sigma_{i}$ and $\Sigma_{ij}$ are 3$\times$3 traceless matrices defined as~\cite{Xin-Li:2023gwh,Becattini:2024uha}
\begin{align}
\Sigma_{1}= & \frac{1}{\sqrt{2}}\left(\begin{array}{ccc}
0 & 1 & 0\\
1 & 0 & 1\\
0 & 1 & 0
\end{array}\right),\nonumber \\
\Sigma_{2}= & \frac{1}{\sqrt{2}}\left(\begin{array}{ccc}
0 & -i & 0\\
i & 0 & -i\\
0 & i & 0
\end{array}\right),\nonumber \\
\Sigma_{3}= & \left(\begin{array}{ccc}
1 & 0 & 0\\
0 & 0 & 0\\
0 & 0 & -1
\end{array}\right),\nonumber \\
\Sigma_{ij}= & \frac{1}{2}(\Sigma_{i}\Sigma_{j}+\Sigma_{j}\Sigma_{i})-\frac{2}{3}\delta_{ij}.
\label{eq:sigma-matrix}
\end{align}

The polarization vector $\vec{\varepsilon}_m$ in the rest frame of the spin-1 particle 
in the spin state $|1m\rangle$ can be defined as 
\begin{align}
&\vec{\varepsilon}_{m=0}=\hat{z}, \\
&\vec{\varepsilon}_{m=\pm 1}=\mp \frac{1}{\sqrt{2}}(\hat{x} \pm i \hat{y}),
\end{align}
where the spin quantization direction is assumed to be $\hat{z}$ direction.  
%\textcolor{red}
{For the measurement of global spin alignment, the spin quantization direction is taken as the direction of global orbital angular momentum, i.e., $\vec{\varepsilon}_{m=0}=\hat{y}$, while other polarization vectors are determined by conditions $\vec{\varepsilon}_{m}^{\,*}\cdot\vec{\varepsilon}_{m^\prime}=\delta_{mm^\prime}$ and $\vec{\varepsilon}_{m=\pm 1}^{\,*}=-\vec{\varepsilon}_{m=\mp 1}$. Another widely used choice is the direction of meson's three-momentum in the laboratory frame, $\vec{\varepsilon}_{m=0}=\vec{p}/|\vec{p}|$, corresponding to the helicity polarization. The spin density matrix for different choices of spin quantization directions are related by the following transformation in the spin space,}
\begin{equation}
\hat{\rho}_V= R\,\hat{\rho}^{\prime}_V\,R^\dagger,
\end{equation}
%\textcolor{red}
{where $\hat{\rho}_V$ and $\hat{\rho}^{\prime}_V$ are two spin density matrices corresponding to two set of polarization vectors $\vec{\varepsilon}_{m=0,\pm 1}$ and $\vec{\varepsilon}^{\,\prime}_{m=0,\pm 1}$, respectively. The transformation matrix $R$ is a $3\times3$ matrix in the spin space with elements given by the inner product of polarization vectors, $R_{mm^\prime}\equiv \vec{\varepsilon}^{\,*}_{m}\cdot\vec{\varepsilon}^{\,\prime}_{m^\prime}$.}

The spin vector $\vec{S}$ is related to the polarization vector $\vec\varepsilon$ by 
$\vec S={\rm Im}({\vec\varepsilon_m}^{\,*}\times\vec\varepsilon_m)$.
If the vector meson is in the pure spin state with $m=0$, the spin vector is vanishing, 
$\vec{S}=\vec\varepsilon_0\times\vec\varepsilon_0=0$.
If the vector meson is in the pure spin state with $m=\pm 1$ we have 
$\vec S={\rm Im}({\vec\varepsilon_{\pm 1}}^{\,*}\times\vec\varepsilon_{\pm 1})=\pm \hat{z}$, 
which is parallel or anti-parallel to the spin quantization direction.

\subsection{Measurement of spin density matrix through angular distribution of decay product}

In high energy reactions, the spin polarization of a particle $A$ is mainly measured through 
the angular distribution of its decay product in two-body decay $A\to 1+2$ in the rest frame of $A$.
The corresponding formulae are derived using symmetry properties and conservation laws in the decay process.

In the rest frame of $A$, the momenta of the particle $1$ and $2$ are denoted as $\vec p=\vec p_1=-\vec p_2$, 
and their spin states are labeled by their helicities $\lambda_1$ and $\lambda_2$ respectively. 
The decay amplitude is given by, 
\begin{align}
A_m(\vec p;\lambda_1\lambda_2)=&\langle\vec p; \lambda_1\lambda_2|\hat U|j_Am_A\rangle,
\end{align}
where $\hat U$ stands for the transition operator, $|j_Am_A\rangle$ denotes the spin state of $A$, and $|\vec{p}; \lambda_1\lambda_2\rangle$ is the helicity state of decay daughters.

We can insert the completeness identity for the helicity states of the two-particle system $|E;jm;\lambda_1\lambda_2\rangle$ with fixed $E$ (the energy of the system), $j$ (total angular momentum quantum number) and $m$ (the eigenvalue of $j_z$). From energy and angular momentum conservation laws in the decay process in $A$'s rest frame, we have $E=M_A$ (the mass of $A$), $j=j_A$ and $m=m_A$. So we obtain, 
\begin{align}
A_m(\vec p;\lambda_1\lambda_2)=&\langle\vec{p}; \lambda_1\lambda_2|M_A;j_Am_A;\lambda_1\lambda_2\rangle \nonumber\\
&\times \langle M_A;j_Am_A;\lambda_1\lambda_2|\hat U|j_Am_A\rangle.
\end{align}
The space rotation invariance demands that the helicity amplitude
\begin{align}
\langle M_A;j_Am_A;\lambda_1\lambda_2|\hat U|j_Am_A\rangle \equiv H_{A}(\lambda_1,\lambda_2),
\end{align}
is independent of $m_A$. So we obtain,
\begin{align}
&A_m(\vec{p};\lambda_1\lambda_2)\nonumber\\
&=\langle\vec{p}; \lambda_1\lambda_2|M_A;j_Am_A;\lambda_1\lambda_2\rangle 
H_{A}(\lambda_1,\lambda_2).
\end{align}
We see that the angular dependence is solely given by the inner product 
$\langle\vec p; \lambda_1\lambda_2|M_A;j_Am_A;\lambda_1,\lambda_2\rangle$,  
which we calculate as follows.
First, we calculate it for $\vec{p}$ in $z$-direction, i.e., $\langle p,0,0; \lambda_1\lambda_2|M_A;j_Am_A;\lambda_1\lambda_2\rangle$, which gives a constant $\sqrt{(2j_A+1)/4\pi}$ independent of $m_A$. 
We then transform $| p,0,0; \lambda_1\lambda_2\rangle$ to $| p,\theta,\varphi; \lambda_1\lambda_2\rangle$ 
in the direction of $\vec{p}$ by a spatial rotation. 
The rotation can be achieved by three successive rotations in Euler angles $(\alpha,\beta,\gamma)=(\varphi,\theta,-\varphi)$ that rotate $\vec p=(p,0,0)$ to $\vec p=(p,\theta,\varphi)$ 
with the rotation operator \cite{Jacob:1959at} 
\begin{equation}
\hat{R}(\varphi,\theta,-\varphi) 
=e^{-i\varphi\hat J_z}e^{-i\theta\hat J_y}e^{i\varphi\hat J_z}.
\label{rotation}
\end{equation}
Hence, we obtain the inner product as 
\begin{align}
&\langle\vec p; \lambda_1\lambda_2|M_A;j_Am_A;\lambda_1\lambda_2\rangle \nonumber\\
&=\sqrt{\frac{2j_A+1}{4\pi}} d_{m_A \lambda}^{j_A*}(\theta) e^{i(m_A-\lambda)\varphi} ,
\end{align}
where $d_{mm'}^{j}(\theta)=\langle jm| e^{-i\theta\hat J_y}|jm'\rangle$ is 
the element of the Wigner rotation matrix and $\lambda=\lambda_1-\lambda_2$.
We note that the definition of the rotation operator $\hat{R}$ in (\ref{rotation}) 
introduces an additional rotation $\gamma=-\varphi$ relative to that in Ref. \cite{Chung:1971ri}.

The spin density matrix of the system of particles 1 and 2 can be defined as
\begin{align}
\hat\rho_{12}=\hat U\hat\rho^A\hat U^\dag,
\end{align}
where $\rho^A$ is the spin density matrix of A. Then the angular distribution is given by
\begin{align}
W(\theta,\varphi)=&N\sum_{\lambda_1,\lambda_2}\langle \vec p;\lambda_1,\lambda_2|\hat\rho_{12}|\vec p;\lambda_1,\lambda_2\rangle\nonumber\\
=&N\sum_{\lambda_1,\lambda_2;m_A,m'_A} |H_A(\lambda_1,\lambda_2)|^2  \rho^A_{m_Am'_A} \nonumber\\
&\times e^{i(m_A-m'_A)\varphi} d_{m_A\lambda}^{j_A*}(\theta) d_{m'_A\lambda}^{j_A}(\theta),
\end{align}
where $\rho^A_{m_Am'_A}=\langle m_A|\hat\rho^A|m'_A\rangle$ denotes elements of $\hat\rho^A$ and $N$ is the normalization constant.

Since the helicity amplitude $H_A(\lambda_1,\lambda_2)$ is generally unknown, 
only in some special cases one can use $W(\theta,\varphi)$ to determine elements of $\hat\rho^A$. 
Here are three such cases. 

(1)  $j_A=1/2$, $\lambda_1=\pm 1/2$, $\lambda_2=0$, 
as in a spin-1/2 hyperon's decay into a spin-1/2 baryon and a pion, $H\to B\pi$, where $H$ denotes a hyperon.
In this case, we have two independent helicity amplitudes $H_A(\pm 1/2,0)$ and 
$W(\theta,\varphi)$ is given by 
\begin{align}
W(\theta,\varphi)%=& N\sum_{\lambda_1;m_A,m'_A}  |H_A(\lambda_1,0)|^2 \rho^A_{m_Am'_A} \nonumber\\
%&\times e^{i(m_A-m'_A)\varphi} d_{m_A\lambda_1}^{1/2*}(\theta) d_{m'_A\lambda_1}^{1/2}(\theta) \nonumber\\
=&\frac{1}{4\pi} \left(1+ \alpha_A\vec P_A\cdot \vec n\right), \label{eq:decHyp}
\end{align}
where $\vec P_A={\rm Tr}(\vec\sigma\hat\rho_A)$ is the polarization vector of $A$,
$\vec n=\vec p/|\vec p~|$ is the momentum direction and
\begin{equation}
\alpha_A=\frac{H_A(1/2,0)-H_A(-1/2,0)}{H_A(1/2,0)+H_A(-1/2,0)},
\end{equation}
is the decay parameter for $A\to 1+2$.
We also see that if parity is conserved in the decay process 
so that $H_A(1/2,0)=H_A(-1/2,0)$, we then have $\alpha_A=0$ and 
the isotropic distribution $W(\theta,\varphi)={1}/{4\pi}$. 
So Eq.~(\ref{eq:decHyp}) can be used to determine $\vec P_A$ only in weak decays.

(2) $j_A=1$, $\lambda_1=\lambda_2=0$, such as in the vector meson's decay into two pions, $V\to \pi\pi$.
In this case, the helicity amplitude $H_A(0,0)$ is a trivial constant 
that can be absorbed into the normalization constant, so that
\begin{align}
W(\theta,\varphi)=&\frac{3}{8\pi} \Bigl\{ (1-\rho^A_{00})+(3\rho^A_{00}-1)\cos^2\theta \nonumber\\
&-{2}\sin^2\theta\bigl[\cos2\varphi {\rm Re}\rho^A_{1-1}-\sin2\varphi{\rm Im}\rho^A_{1-1}\bigr] \nonumber\\
&-\sqrt{2}\sin2\theta\bigl[\cos\varphi({\rm Re}\rho^A_{10}-{\rm Re}\rho^A_{-10}) \nonumber\\
&-\sin\varphi({\rm Im}\rho^A_{10}+{\rm Im}\rho^A_{-10})\bigr]\Bigr\}, \label{eq:decVpp}
\end{align}
If we integrate over $\varphi$, we obtain,
\begin{align}
W(\theta)=\frac{3}{4} &\Bigl\{ (1-\rho^A_{00})+(3\rho^A_{00}-1)\cos^2\theta\Bigr\}. 
\label{eq:decVppint}
\end{align}
We see that by measuring the distribution in $\theta$ one can extract the value of $\rho^A_{00}$.

(3) $j_A=1$, $\lambda_1=\pm 1/2, \lambda_2=\pm 1/2$, such as in the vector meson's decay into 
a dilepton pair, $V\to e^+e^-$.
Here, we have four combinations of $\lambda_1$ and $\lambda_2$.
In this case, only if parity and helicity conservation are valid so that 
$H_A(-\lambda_1,-\lambda_2)=H_A(\lambda_1,\lambda_2)$
and $H_A(\lambda_1,\lambda_2)\not= 0$ only for $\lambda_1=-\lambda_2$,
only one non-vanishing helicity amplitude is left and can be absorbed into the normalization constant.
We then obtain the angular distribution as 
\begin{align}
W(\theta,\varphi)=&\frac{3}{8\pi(1+\rho^A_{00})} \Bigl\{ 1+\lambda_\theta\cos^2\theta \nonumber\\
&+\lambda_\varphi\sin^2\theta\cos2\varphi +\lambda_{\theta\varphi}\sin2\theta\sin2\varphi \nonumber\\
&+\lambda^\perp_\varphi \sin^2\theta\sin2\varphi+\lambda^\perp_{\theta\varphi}\sin2\theta\sin\varphi\Bigr\}, \label{eq:decVll2}
\end{align}
where the $\lambda$ coefficients are related to elements of $\rho^A$ by
\begin{align}
&\lambda_\theta=\frac{1-3\rho^A_{00}}{1+\rho^A_{00}}, \\
&\lambda_\varphi=\frac{2{\rm Re}\rho^A_{1-1}}{1+\rho^A_{00}}, \\
&\lambda_{\theta\varphi}=\frac{\sqrt{2}{\rm Re}(\rho^A_{10}-\rho^A_{-10})}{1+\rho^A_{00}}, \\
&\lambda_\varphi^\perp=-\frac{2{\rm Im}\rho^A_{1-1}}{1+\rho^A_{00}}, \\
&\lambda_{\theta\varphi}^\perp=-\frac{\sqrt{2}{\rm Im}(\rho^A_{10}+\rho^A_{-10})}{1+\rho^A_{00}}.
\end{align}
In principle, one can extract elements of $\rho^A$ from $W(\theta,\varphi)$ through these coefficients.

Measurements on hyperon polarization and vector meson's spin alignment 
have been carried out in different high energy reactions using Eqs.~(\ref{eq:decHyp}-\ref{eq:decVll2})
(see e.g. Refs. \cite{Lesnik:1975my,Bunce:1976yb,Bensinger:1983vc,Gourlay:1986mf,TASSO:1984nda,ALEPH:1996oew,OPAL:1997oem},
\cite{DELPHI:1997ruo,OPAL:1997vmw,OPAL:1997nwj,OPAL:1999hxs,NOMAD:2006kuc} and \cite{Chen:2007zzq,Selyuzhenkov:2007ab,STAR:2008lcm,STAR:2007ccu}).
In this brief review, we concentrate on spin alignment of vector mesons with light flavors 
in comparison with hyperon polarization, thus only Eqs.~(\ref{eq:decHyp}) and (\ref{eq:decVpp}) are involved.

\section{Experimental results}
\subsection{Vector meson's spin alignments in $e^+e^-$-annihilations}

\begin{figure}[!ht]
\includegraphics[width=0.45\textwidth]{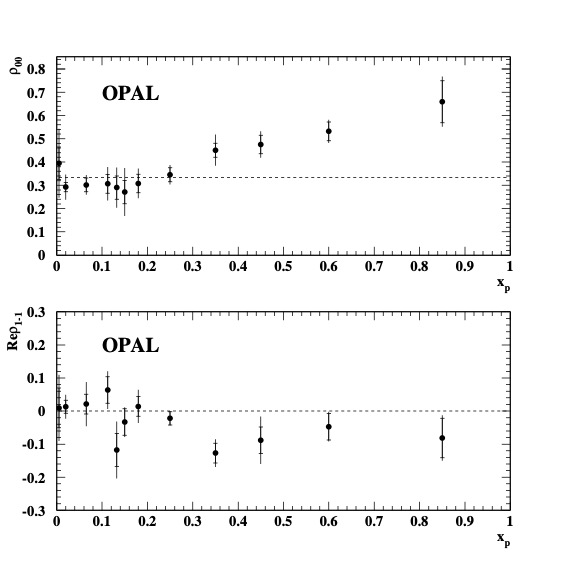}
\caption{Experimental data obtained by OPAL Collaboration at LEP 
for the spin alignment $\rho_{00}$ and off-diagonal element $\rho_{1-1}$ of $K^{*0}$. 
This figure is taken from Ref.~\cite{OPAL:1997vmw}. } 
\label{fig:opal97}
\end{figure}

The earliest measurements of vector meson's spin alignments in high energy reactions might be made 
at the Large Electron-Position collider (LEP) at European Organization for Nuclear Research (CERN) in the 1990s~\cite{DELPHI:1997ruo,OPAL:1997vmw,OPAL:1997nwj,OPAL:1999hxs}, 
where one of the popular polarization axis was the helicity axis defined by the individual particle's momentum direction. 
Measurements have been carried out by DELPHI and OPAL Collaborations for $K^{*0}$ and $\phi$ mesons and even for heavy flavor meson such as $B$ 
and $D^*$~\cite{DELPHI:1997ruo,OPAL:1997vmw,OPAL:1997nwj,OPAL:1999hxs}. 
They also measured the off-diagonal element such as $\rho_{1,-1}$. 
As an example, we show the results from Ref.~\cite{OPAL:1997vmw} 
in Fig.~\ref{fig:opal97}. 

From these measurements~\cite{DELPHI:1997ruo,OPAL:1997vmw,OPAL:1997nwj,OPAL:1999hxs} in $e^+ e^-$ collisions, we see clearly that $\rho_{00}$ is significantly larger than 1/3 in the fragmentation region $x_p>0.3$ ($x_p \equiv p/p_{\rm beam}$) which show that the spin of vector mesons is significantly aligned in that region. 
However, at small fractional momenta $x_p \le 0.3$, null results for $K^{*0}$ and $\phi$ mesons were reported in $e^+e^-$ collisions~\cite{DELPHI:1997ruo,OPAL:1997vmw}.

The data have attaracted much theoretical attention~\cite{Anselmino:1997ui,Anselmino:1998jv,Anselmino:1999cg,Xu:2001hz,Xu:2003fq,Chen:2016moq,Chen:2016iey,Chen:2020pty} and we will come back to this later in Sec.~\ref{sec:FFs}.

\subsection{Global spin alignments of vector mesons in heavy-ion collisions}

New measurements of the vector meson's spin alignment in heavy-ion collisions are different from conventional studies, where a new polarization axis along the nucleus-nucleus system's orbital angular momentum is defined~\cite{Liang:2004ph,Liang:2004xn}, and this is the so-called global spin alignment. In experiment, the quantization axis is determined by the normal of the reaction plane, which can be reconstructed by using the charge particle momentum distribution collected in the detector~\cite{STAR:2022fan}. Particles of interest, for example, the vector-meson $\phi$ and $K^{*0}$ are observed by paring of their decay daughters ($K^{\pm}$, $\pi^{\pm}$) with subtraction of the combinatorial background. Then the polar angle distribution of Eq.~(\ref{eq:decVppint}) is analyzed, and the $\rho_{00}$ is extracted after correction for detection efficiency and acceptance~\cite{STAR:2022fan}.

\begin{figure}[!ht]
\includegraphics[width=0.45\textwidth]{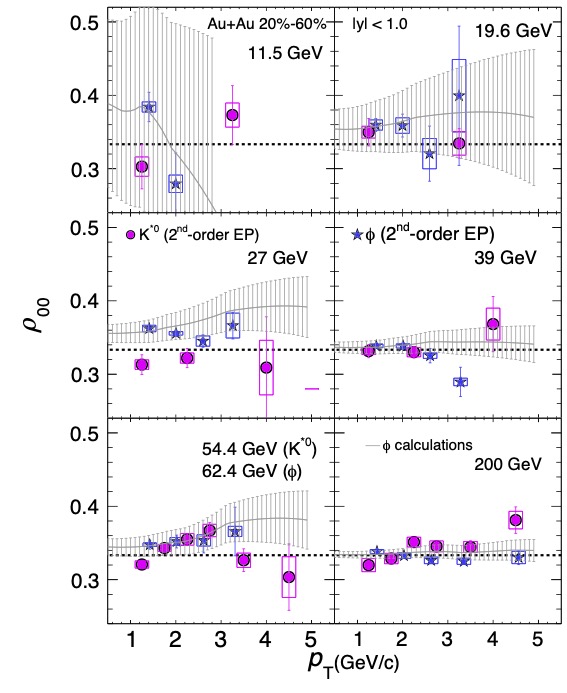}
\caption{Measurements of the vector meson's spin density matrix element $\rho_{00}$ as functions of transverse momentum ($p_{\rm T}$) for $K^{*0}$ and $\phi$ mesons in Au+Au collisions at different energies. Data points are STAR's measurements~\cite{STAR:2022fan}, bands are theoretical calculations for $\phi$ mesons~\cite{Sheng:2022wsy}.} 
\label{fig:rho00pt}
\end{figure}

In 2008 the STAR Collaboration performed a measurement of the vector meson's global spin alignment in Au+Au collisions at $\sqrt{s_{\rm NN}}$ = 200 GeV. Due to limited statistics and only covered top RHIC collision energies, no significant results were reported~\cite{STAR:2008lcm}. Since 2010, the collaboration has been collecting and analyzing data of higher statistics and at lower collision energies, including data in the Beam Energy Scan Phase I (BES-I) runs and high statistics Au+Au runs at $\sqrt{s_{\rm NN}}$ = 200 GeV. The analysis was focused on mid-central collisions (20-60\%) where larger system angular momenta are expected in comparison with the values in central or peripheral collisions. Fig.~\ref{fig:rho00pt} presents the transverse momentum dependence of $\rho_{00}$ for $K^{*0}$ and $\phi$ in 20-60\% central Au+Au collisions at $\sqrt{s_{\rm NN}}$ = 11.5, 19.6, 27, 39, 62.4 (54.4), 200 GeV. The $\rho_{00}$ results show a non-trivial $p_{\rm T}$ dependence. For $K^{*0}$ mesons at $\sqrt{s_{\rm NN}}$ = 54.4 and 200 GeV, they are larger than 1/3 with about $2\sigma$ significance at intermediate $p_{\rm T}$. At low beam energies the statistics is not sufficient to observe any significant deviation from 1/3. From the calculation in Ref.~\cite{Liang:2004xn}, one naively expect $\rho_{00}$ to be smaller than 1/3 due to hadronization of the polarized quark and antiquark via quark combination, and larger than 1/3 due to fragmentation of the quark and antiquark. On the other hand, for $\phi$ mesons for all energies considered, we see that the departure of $\rho_{00}$ from 1/3 mainly occurs at $p_{\rm T}\sim$ 1.0 - 2.4 GeV/c, while at higher $p_{\rm T}$ the result can be regarded as being consistent with 1/3 within $\sim 2\sigma$ or less significance.

\begin{figure}
\includegraphics[width=7.0cm]{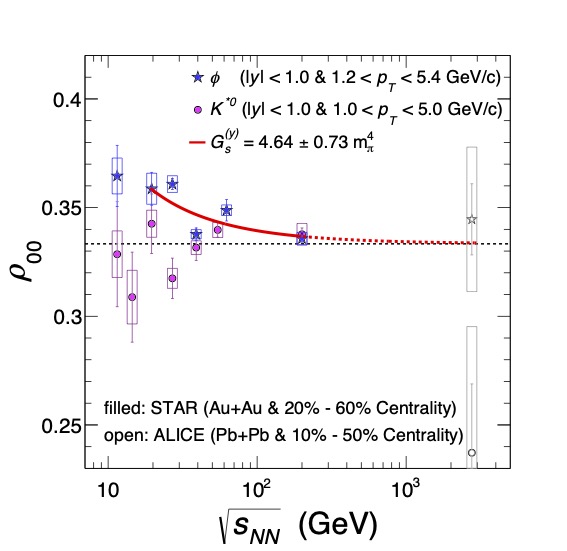}
\caption{\label{fig:rho00_data}Measurements of $\rho_{00}$ with respect to the system orbital angular momentum in high energy heavy-ion collisions.
Stars represent the data for $\phi$ mesons~\cite{STAR:2022fan,ALICE:2019aid}, circles represent the data for $K^{*0}$ mesons~\cite{STAR:2022fan,ALICE:2019aid}.
The solid red line is the prediction from a non-relativistic coalescence model with strong force fields~\cite{Sheng:2019kmk}. 
More sophisticated and complete calculations using the relativistic coalescence model and spin kinetic equation~\cite{Sheng:2022ffb}
were presented in Ref.~\cite{Sheng:2022wsy,Sheng:2023urn}.}
\label{fig:rho00energy}
\end{figure}

The STAR collaboration also studied the collision energy dependence by integrating $\rho_{00}(p_{\rm T})$ with the weight 1/(stat. error)$^2$. Fig.~\ref{fig:rho00energy} presents $\rho_{00}$ for $K^{*0}$ and $\phi$ mesons in 20-60\% central Au+Au collisions at collision energies ranging from $\sqrt{s_{\rm NN}}$ = 11.5 to 200 GeV~\cite{STAR:2022fan}. We see that $\rho_{00}$ for $\phi$ mesons increases with decreasing the collision energy, while $\rho_{00}$ for $K^{*0}$ mesons fluctuates around 1/3 with the collision energy. To quantify the effect, the average of $\rho_{00}$ is taken over lower collision energies for $K^{*0}$ and $\phi$. The $\rho_{00}$ for $\phi$ mesons, averaged over beam energies between 11.5 and 62.4 GeV, is 0.3512 $\pm$ 0.0017 (stat.) $\pm$ 0.0017 (syst.). Taking the total uncertainties as the sum in quadrature of statistical and systematic uncertainties, the result indicates that $\rho_{00}$ for $\phi$ mesons is above 1/3 with a significance of $7.4\sigma$, representing the first observation of the global spin alignment. The $\rho_{00}$ for $K^{*0}$ mesons, averaged over beam energies between 11.5 and 54.4 GeV, is 0.3356 $\pm$ 0.0034 (stat.) $\pm$ 0.0043 (syst.) and is consistent with 1/3. The measurements of the ALICE collaboration in Pb+Pb collisions at 2.76 TeV~\cite{ALICE:2019aid}, taken from the closest data points to the mean $p_{\rm T}$ for the $p_{\rm T}$ range used in STAR's measurements, are also shown for comparison in Fig.~\ref{fig:rho00energy}. The $\rho_{00}$ data point for $K^{*0}$ and $\phi$ mesons from ALICE collaboration is more or less consistent with 1/3 with large uncertainties.

According to various studies, there are many sources that contribute to the global spin alignment of $\phi$ mesons including vortical flows \cite{Yang:2017sdk},  electromagnetic fields \cite{Yang:2017sdk} generated by the electric currents carried by the colliding nuclei, local spin alignment \cite{Xia:2020tyd}, quark polarization along the direction of its momentum (helicity polarization) \cite{Gao:2021rom}, the spin alignment by fragmentation \cite{Liang:2004xn} of polarized quarks, and the shear stress tensor \cite{Li:2022vmb,Wagner:2022gza,Dong:2023cng}. However, these conventional mechanisms are not sufficient to account for the observed $\rho_{00}$ for $\phi$ mesons.  
It was also proposed that local correlations or fluctuations in turbulent
color fields \cite{Muller:2021hpe} and glasma fields~\cite{Kumar:2023ghs}
can also generate a significant contribution to $\rho_{00}$. 
A recent theoretical development based on local correlations or fluctuations of $\phi$ vector fields can describe the experimental data. The $\phi$ vector field is the '33' component of the SU(3) vector multiplet induced by currents of pseudo-Goldstone bosons~\cite{Manohar:1983md}, and it can polarize $s$ and $\bar{s}$ quarks in the same way as the electromagnetic field does. The solid curve of Fig.~\ref{fig:rho00energy} is such a fit to the data. We see that the calculation describes the data reasonably well. The correlation of the $\phi$ vector field can be quantified by $\langle{\vec E}_\phi^2\rangle$ and $\langle{\vec B}_\phi^2\rangle$ and can be extracted by fitting data, where ${\vec E}_\phi$ and ${\vec B}_\phi$ are electric and magnetic components of the $\phi$ vector field respectively. We will discuss more details about the theoretical model in Sec.~\ref{rel-qcm}.

\section{Vector meson's spin alignments in quark combination models}

It was well-known that huge angular momenta are generated in non-central high energy heavy-ion collisions~\cite{Liang:2004ph,Gao:2007bc}. Due to spin-orbit couplings in QCD interaction, such huge angular momenta lead to the global polarization in quark scatterings in the form of the hyperon polarization~\cite{Liang:2004ph,Gao:2007bc} and vector meson's spin alignment~\cite{Liang:2004xn}, the so-called GPE.

The GPE depends not only on the global quark polarization but also on the hadronization mechanism. 
In relativistic heavy-ion collisions, it is envisaged that a system of decomfined quarks and anti-quarks 
is created in the central rapidity and moderate transverse momentum region.
Various aspects of experimental data suggest that hadronization of the system proceeds via combination 
of quarks and anti-quarks, a mechanism phrased as ``quark recombination'', ``quark coalescence'' 
or simply as ``quark combination''. 
We simply refer to it as ``quark combination model'' or ``quark coalescence model'' in this paper. 
We will summarize the results of vector meson's spin alignments by the quark combination model in this section.

\subsection{Vector meson's spin alignments in non-relativistic quark combination models}

In a non-relativistic quark combination model, the physics in the process that $q$ (quark) and $\bar{q}$ (antiquark) 
combine to $M$ (meson) can be demonstrated in a clear way.
Here, it is envisaged that in the combination process the vector meson's spin 
is just the sum over the quark's and anti-quark's spins.
Hence the spin density matrix and the spin alignment of the vector meson can be calculated from 
that of the quark and anti-quark. 
Such a calculation is straightforward and was carried out in Ref. \cite{Liang:2004xn}, which we will 
summarize in this subsection.

The global quark polarization was taken as a constant so that the spin density matrix for the quark and antiquark takes the diagonal form~\cite{Liang:2004xn}
\begin{align}
 \hat\rho^q=\frac{1}{2}
 \left(\begin{array}{cc}
1+P_q & 0  \\
0 & 1-P_q
\end{array} \right). \label{eqrhoq}
\end{align}
The spin density matrix of the $q_1\bar{q}_2$ system was taken as a direct product of the quark's and anti-quark's,
\begin{align}
 \hat\rho^{q_1\bar q_2}=\hat\rho^{q_1}\otimes\hat\rho^{\bar q_2}.
\end{align}
The elements of the spin density matrix $\hat\rho^{V}$ for the vector meson is obtained from $\hat\rho^{q_1\bar q_2}$ as 
\begin{align}
 \rho^V_{m'm}=\langle jm'| \hat{\rho}^{q_1\bar q_2}|jm\rangle,
\end{align}
which leads to
\begin{align}
\rho^V_{m'm}=&\sum_{m_1,m_2,m_1',m_2'}\rho^{q_1\bar q_2}_{m_1'm_2',m_1m_2}\nonumber\\
&\times\langle j_V m'|m'_1m'_2\rangle \langle m_1m_2|j_V m\rangle,
\end{align}
where $|j_V m\rangle$ is the vector meson's spin state in the constituent quark model with 
$j_V=1$ and $m=0,\pm1$, and $\langle j_V m|m_1 m_2\rangle$ are Clebsch-Gordan coefficients. 
After a straightforward calculation, we obtain the normalized spin alignment $\rho^V_{00}$ as~\cite{Liang:2004xn},
\begin{equation}
\rho^V_{00}=\frac{1-P_{q_1}P_{\bar q_2}}{3+P_{q_1}P_{\bar q_2}} . \label{eqrhoV00}
\end{equation}
If we take $P_{q_1}=P_{\bar q_2}=P_q$ (flavor blind for quarks and antiquarks), we simply obtain 
\begin{equation}
\rho^V_{00}=\frac{1-P_{q}^2}{3+P_{q}^2}. \label{eqrhoV00s}
\end{equation}
In exact the same way, we obtain the global hyperon polarization $P_H=P_q$ for $\Lambda$, $\Sigma$, and $\Xi$~\cite{Liang:2004ph}.

From Eqs.~(\ref{eqrhoV00}) and (\ref{eqrhoV00s}), we see that, in contrast to $P_H$, 
$\rho^V_{00}$ is quadratic in $P_q$ and should be less than $1/3$. 
We emphasize that the case considered in Ref. \cite{Liang:2004xn} is over simplified 
in the fact that only the spin degree of freedom for the quark and antiquark is considered, 
neglecting other degree(s) of freedom and the correlation among the quark's and antiquark's polarization.
So it might not be a surprise that the STAR data~\cite{STAR:2017ckg,STAR:2022fan}
for the global spin alignment of $\phi$ mesons show a large deviation from 1/3, far beyond 
the value estimated from $P_\Lambda$'s data by its square. 
In fact, if we make a step forward by considering the dependence of $P_q$ on other degree of freedom~\cite{Sheng:2019kmk,Sheng:2020ghv,Sheng:2022ffb,Sheng:2022wsy}, 
$P_{q_1}P_{\bar q_2}$ in Eq.~(\ref{eqrhoV00}) should be replaced by $\langle P_{q_1}P_{\bar q_2}\rangle$, so we have  
\begin{equation}
\rho^V_{00}=\frac{1-\langle P_{q_1}P_{\bar q_2}\rangle}{3+\langle P_{q_1}P_{\bar q_2}\rangle}. \label{eqrhoV00m}
\end{equation}
The STAR data~\cite{STAR:2017ckg,STAR:2022fan} indicate 
\begin{equation}
\langle P_{q_1}P_{\bar q_2}\rangle\not=\langle P_{q_1}\rangle\langle P_{\bar q_2}\rangle. \label{eqPq12}
\end{equation}
This means that there should be a strong correlation between the polarization of the quark and antiquark. 
Hence, the study of the global spin alignment for the vector meson in heavy-ion collisions 
provides a unique opportunity for exploring the local correlation in the quark's and antiquark's polarization, a new window for the study of properties of QGP.

It is also emphasized in Ref. \cite{Liang:2023talk} that the average in Eq.~(\ref{eqrhoV00m}) is two-folded: 
the first average is taken inside the vector meson's wave function (local correlation) 
and the second average is taken in the range outside the vector meson's wave function (long-range correlation). More measurements are needed to tell the difference between two types of correlation~\cite{Liang:2023talk}.
A systematic formulation has been presented in~\cite{Lv:2024uev}. It has bee extended to spin 3/2 baryons in \cite{Zhang:2024hyq}.

\subsection{Vector meson's spin alignments in quark combination models with phase space dependence}\label{rel-qcm}

\subsubsection{Non-relativistic model}

In this subsection, we consider the non-relativistic quark combination
model with phase space dependence \cite{Sheng:2020ghv,Xin-Li:2023gwh}. We extend the density operator
(\ref{eq:rho1/2}) to the phase space as
\begin{align}
\hat{\rho}^{q}= & \sum_{rs}\int[d^{3}{\vec p}][d^{3}{\vec q}]\int d^{3}{\vec x}e^{-i{\vec q}\cdot{\vec x}}\nonumber \\
 & \times w_{rs}^{q}({\vec x},{\vec p})\left|r,{\vec p}+\frac{{\vec q}}{2}\right\rangle \left\langle s,{\vec p}-\frac{{\vec q}}{2}\right|\,,
\end{align}
where $r,s=\pm$ denote spin states, $\left|r,{\vec p}\right\rangle $
denotes the spin-momentum state, and the momentum measure is defined
as $[d^{3}{\vec p}]\equiv d^{3}{\vec p}/(2\pi)^{3}$. The weight function
$w_{rs}^{q}({\vec x},{\vec p})$ is the Wigner function in the spin-phase
space and can be obtained as
\begin{equation}
w_{rs}^{q}({\vec x},{\vec p})=\int[d^{3}{\vec q}]e^{i{\vec q}\cdot{\vec x}}\left\langle r,{\vec p}+\frac{{\vec q}}{2}\right|\hat{\rho}^{q}\left|s,{\vec p}-\frac{{\vec q}}{2}\right\rangle \,.\label{eq:quark-Wigner}
\end{equation}
It is a $2\times2$ Hermitian matrix in the spin space and can be parameterized as
\begin{equation}
w_{rs}^{q}({\vec x},{\vec p})=\frac{1}{2}f_{q}({\vec x},{\vec p})\left[\delta_{rs}+{\vec \sigma}_{rs}\cdot{\vec P}_{q}({\vec x},{\vec p})\right]\,.
\end{equation}
Here $f_{q}({\vec x},{\vec p})$ is the un-polarized distribution function,
and ${\vec P}_{q}({\vec x},{\vec p})$ is the spin polarization in phase
space. The density operator for the anti-quark can be defined in the
same way. The density operator for a quark-antiquark pair $q_{1}\overline{q}_{2}$
is defined from the direct product of quark's and anti-quark's one
as $\hat{\rho}^{q_{1}\bar{q}_{2}}\equiv\hat{\rho}^{q_{1}}\otimes\hat{\rho}^{\bar{q}_{2}}$.
The Wigner function can be normalized as
\begin{equation}
\sum_{s}\int d^{3}{\vec x}\int[d^{3}{\vec p}]w_{ss}({\vec x},{\vec p})=1.
\end{equation}
The quark's spin quantization direction can be any direction, 
e.g., the $z$-direction, without loss of generality. 
 
The elements of the vector meson's density matrix are obtained as
\begin{eqnarray}
&&\rho_{m^{\prime}m}^{V}({\vec x},{\vec p}) \nonumber\\ 
& = & \frac{1}{N({\vec x},{\vec p})}\int[d^{3}{\vec q}]e^{i{\vec q}\cdot{\vec x}}\nonumber \\
 &  & \times\left\langle j_{V}m^{\prime};{\vec p}+\frac{{\vec q}}{2}\right|\hat{\rho}^{q_{1}}\otimes\hat{\rho}^{\bar{q}_{2}}\left|j_{V}m;{\vec p}-\frac{{\vec q}}{2}\right\rangle \nonumber \\
 & = & \frac{1}{N({\vec x},{\vec p})}\int d^{3}{\vec x}_{b}[d^{3}{\vec p}_{b}][d^{3}{\vec q}_{b}]\exp(-i{\vec q}_{b}\cdot{\vec x}_{b})\nonumber \\
 &  & \times\varphi_{V}^{\ast}\left({\vec p}_{b}+\frac{{\vec q}_{b}}{2}\right)\varphi_{V}\left({\vec p}_{b}-\frac{{\vec q}_{b}}{2}\right)\nonumber \\
 &  & \times\sum_{r_{1},r_{2},s_{1},s_{2}}w_{m_{1}^{\prime}m_{1}}^{q_{1}}({\vec x}_{1},{\vec p}_{1})w_{m_{2}^{\prime}m_{2}}^{\bar{q}_{2}}({\vec x}_{2},{\vec p}_{2})\nonumber \\
 &  & \times\left\langle j_{V}m^{\prime}\mid m_{1}^{\prime}m_{2}^{\prime}\right\rangle \left\langle m_{1}m_{2}\mid j_{V}m\right\rangle \,,\label{eq:density-vector-meson}
\end{eqnarray}
where ${\vec x}_{1,2}\equiv{\vec x}\pm{\vec x}_{b}/2$ and ${\vec p}_{1,2}\equiv{\vec p}/2\pm{\vec p}_{b}$
are positions and momenta for the quark $q_{1}$ and the antiquark
$\overline{q}_{2}$ respectively, and $N({\vec x},{\vec p})$ is the
normalization factor that ensures $\sum_{m=0,\pm1}\rho_{mm}^{V}({\vec x},{\vec p})=1$.
Since we work in the non-relativistic limit, the spin and momentum
can be decoupled. Therefore Clebsch-Gordan coefficients $\left\langle m_{1}m_{2}\mid j_{V}m\right\rangle $ 
and $\left\langle j_{V}m^{\prime}\mid m_{1}^{\prime}m_{2}^{\prime}\right\rangle$ as well as   
the meson's wave function $\varphi_{V}$ appear in Eq. (\ref{eq:density-vector-meson}).
By further neglecting the phase-space dependence of unpolarized distribution
functions and choosing $\varphi_{V}$ as a Gaussian distribution with
the width $a_{V}$ in the momentum space, we derive the spin alignment
and other parameters in the angular distribution (\ref{eq:decVpp}),
\begin{align}
& \rho_{00}^{V}({\vec x},{\vec p})\approx\frac{1}{3}-\frac{2}{3}\left\langle P_{q_{1}}^{z}P_{\bar{q}_{2}}^{z}\right\rangle _{V}+\frac{2}{9}\left\langle {\vec P}_{q_{1}}\cdot{\vec P}_{\bar{q}_{2}}\right\rangle _{V}\,,\nonumber \\
& \text{Re}\rho_{1,-1}^{V}({\vec x},{\vec p})\approx\frac{1}{3}\left\langle P_{q_{1}}^{x}P_{\bar{q}_{2}}^{x}-P_{q_{1}}^{y}P_{\bar{q}_{2}}^{y}\right\rangle _{V},\nonumber \\
& -\text{Im}\rho_{1,-1}^{V}({\vec x},{\vec p})\approx\frac{1}{3}\left\langle P_{q_{1}}^{x}P_{\bar{q}_{2}}^{y}+P_{q_{1}}^{y}P_{\bar{q}_{2}}^{x}\right\rangle _{V},\nonumber \\
& \text{Re}\left[\rho_{1,0}-\rho_{-1,0}\right]({\vec x},{\vec p})
\approx \frac{\sqrt{2}}{3}\left\langle P_{q_{1}}^{z}P_{\bar{q}_{2}}^{x}+P_{q_{1}}^{x}P_{\bar{q}_{2}}^{z}\right\rangle _{V},\nonumber \\
& -\text{Im}\left[\rho_{1,0}+\rho_{-1,0}\right]({\vec x},{\vec p}) 
\approx \frac{\sqrt{2}}{3}\left\langle P_{q_{1}}^{y}P_{\bar{q}_{2}}^{z}+P_{q_{1}}^{z}P_{\bar{q}_{2}}^{y}\right\rangle _{V}, \nonumber\\
& \label{eq:tensor-polar}
\end{align}
where the correlation between quark and antiquark's polarizations
is defined as
\begin{align}
\left\langle P_{q_{1}}^{i}P_{\bar{q}_{2}}^{j}\right\rangle _{V}\equiv & \frac{1}{\pi^{3}}\int d^{3}{\vec x}_{b}d^{3}{\vec p}_{b}\exp\left(-\frac{{\vec p}_{b}^{2}}{a_{V}^{2}}-a_{V}^{2}{\vec x}_{b}^{2}\right)\nonumber \\
 & \times P_{q_{1}}^{i}({\vec x}_{1},{\vec p}_{1})P_{\bar{q}_{2}}^{j}({\vec x}_{2},{\vec p}_{2})\,.
\end{align}
We note that in Eq. (\ref{eq:tensor-polar}) the spin quantization
direction is set to the $z$-direction.

\subsubsection{Relativistic model in quantum kinetic approach}

For a relativistic quark or antiquark, its polarization four-vector is
always perpendicular to its four-momentum, implying that the spin with momentum 
cannot be decoupled as in the non-relativistic case. 
Consequently, the vector meson's spin cannot be obtained by 
the constituent quark's and anti-quark's spins through the angular momentum coupling 
in non-relativistic quantum mechanics. Based on the Kadanoff-Baym equation in the closed-time-path formalism, a relativistic spin kinetic theory for vector mesons has been constructed \cite{Sheng:2022ffb} to explain experimental data on the spin alignment \cite{STAR:2017ckg}.

In order to describe spin transport phenomena, we use the matrix-valued
spin-dependent distribution (MVSD) $f_{m_{1}m_{2}}^{V}(x,{\vec p})$
for the vector meson. The spin Boltzmann equation with coalescence
and dissociation collision terms reads,
\begin{eqnarray}
 &  & p\cdot\partial_{x}f_{m_{1}m_{2}}^{V}(x,{\vec p})\nonumber \\
 & = & \frac{1}{16}\sum_{m_{1}^{\prime},m_{2}^{\prime}}\left[\epsilon_{\mu}^{\ast}(m_{1},{\vec p})\epsilon_{\nu}(m_{1}^{\prime},{\vec p})\delta_{m_{2}m_{2}^{\prime}}\right.\nonumber \\
 &  & \left.+\delta_{m_{1}m_{1}^{\prime}}\epsilon_{\mu}^{\ast}(m_{2}^{\prime},{\vec p})\epsilon_{\nu}(m_{2},{\vec p})\right]\nonumber \\
 &  & \times\mathcal{C}_{m_{1}^{\prime}m_{2}^{\prime}}^{\mu\nu}(x,{\vec p}),\label{eq:sbe-1}
\end{eqnarray}
where $\epsilon^{\mu}(m,{\vec p})$ denotes the meson's normalized
polarization vector perpendicular to $p^{\mu}$, and $m_{1}$, $m_{2}$,
$m_{1}^{\prime}$, and $m_{2}^{\prime}$ label the vector meson's
spin states along the spin quantization direction with three values
0 and $\pm1$. Here we have assumed that the vector meson is on its
mass-shell $p^{0}=\sqrt{m_{V}^{2}+{\vec p}^{2}}$. The gain and loss
terms in $\mathcal{C}_{m_{1}^{\prime}m_{2}^{\prime}}^{\mu\nu}$ correspond
to the coalescence and dissociation processes for the quark and antiquark,
respectively. In the dilute limit when distribution functions of the
meson, the constituent quark and antiquark are much less than unity,
Eq. (\ref{eq:sbe-1}) is simplified as 
\begin{align}
p\cdot\partial_{x}f_{m_{1}m_{2}}^{V}(x,{\vec p})= & \frac{1}{8}\left[\epsilon_{\mu}^{\ast}(m_{1},{\vec p})\epsilon_{\nu}(m_{2},{\vec p})\mathcal{C}_{\text{coal}}^{\mu\nu}(x,{\vec p})\right.\nonumber \\
 & \left.-\mathcal{C}_{\text{diss}}({\vec p})f_{m_{1}m_{2}}^{V}(x,{\vec p})\right],\label{eq:Boltzmann-eq-new}
\end{align}
where the contribution from the dissociation is proportional to the
MVSD with the coefficient $\mathcal{C}_{\text{diss}}$ being independent
of the MVSD. The coalescence kernel $\mathcal{C}_{\text{coal}}^{\mu\nu}$
can be obtained from the Kadanoff-Baym equation, 
\begin{eqnarray}
\mathcal{C}_{\text{coal}}^{\mu\nu}(x,{\vec p}) & = & \int\frac{d^{3}{\vec p}^{\prime}}{(2\pi)^{2}}\frac{\delta(E_{{\vec p}}^{V}-E_{{\vec p}^{\prime}}^{\bar{q}}-E_{{\vec p}-{\vec p}^{\prime}}^{q})}{E_{{\vec p}^{\prime}}^{\bar{q}}E_{{\vec p}-{\vec p}^{\prime}}^{q}}\nonumber \\
 &  & \times\text{Tr}\left\{ \Gamma^{\nu}(p^{\prime}\cdot\gamma-M_{\bar{q}})\right.\nonumber \\
 &  & \times[1+\gamma_{5}\gamma\cdot P_{\bar{q}}(x,{\vec p}^{\prime})]\nonumber \\
 &  & \times\Gamma^{\mu}[(p-p^{\prime})\cdot\gamma+M_{q}]\nonumber \\
 &  & \times\left.[1+\gamma_{5}\gamma\cdot P_{q}(x,{\vec p}-{\vec p}^{\prime})]\right\} \nonumber \\
 &  & \times f_{\bar{q}}(x,{\vec p}^{\prime})f_{q}(x,{\vec p}-{\vec p}^{\prime})\,,\label{eq:coalescence-kernel}
\end{eqnarray}
where $M_q$ and $M_{\bar{q}}$ are quark and antiquark masses respectively, 
$E_{{\vec p}}^{V}$, $E_{{\vec p}-{\vec p}^{\prime}}^{q}$, and
$E_{{\vec p}^{\prime}}^{\bar{q}}$ are energies for the vector meson,
the constituent quark and antiquark respectively, and $\Gamma^{\mu}$
is the effective vertex of quark-antiquark-meson. Here $f_{q/\bar{q}}$
and $P_{q/\bar{q}}^{\mu}$ are the unpolarized and spin polarization
distributions for the quark/antiquark respectively. By neglecting
the space dependence (the distributions are homogeneous in space)
and setting $f_{m_{1}m_{2}}^{V}=0$ before the hadronization stage
in heavy-ion collisions, one can obtain a formal solution to Eq. (\ref{eq:Boltzmann-eq-new}).
Spin density matrix elements can be derived as \cite{Sheng:2022wsy},
\begin{equation}
\rho_{m_{1}m_{2}}^{V}(x,{\vec p})=\frac{\epsilon_{\mu}^{\ast}(m_{1},{\vec p})\epsilon_{\nu}(m_{2},{\vec p})\mathcal{C}_{\text{coal}}^{\mu\nu}(x,{\vec p})}{\sum_{m=0,\pm1}\epsilon_{\alpha}^{\ast}(m,{\vec p})\epsilon_{\beta}(m,{\vec p})\mathcal{C}_{\text{coal}}^{\alpha\beta}(x,{\vec p})}.
\end{equation}
This is a relativistic formula for the vector meson's spin density
matrix built from the coalescence kernel (\ref{eq:coalescence-kernel})
encoding the polarization distributions of the constituent quark and
antiquark. Clearly the vector meson's spin density matrix has non-trivial
dependence on the spin polarization and momenta of the quark and antiquark.

We consider that $s$ and $\bar{s}$ are polarized in a thermal medium
by an effective vector field called the $\phi$ field \cite{Sheng:2019kmk,Sheng:2022wsy,Sheng:2022ffb}
induced by currents of pseudo-Goldstone bosons \cite{Manohar:1983md}.
The spin polarization of $s$ and $\bar{s}$ is then given by
\begin{equation}
P_{q/\bar{q}}^{\mu}(x,{\vec p})\approx\pm\frac{g_{\phi}}{4M_{s}(u\cdot p)T_{\text{h}}}\epsilon^{\mu\nu\alpha\beta}p_{\nu}F_{\alpha\beta}^{\phi}(x),\label{eq:polar-vector}
\end{equation}
where $u^{\mu}$ denotes the velocity of the thermal background at
the effective hadronization temperature $T_{\text{h}}$, $g_{\phi}$
is the effective coupling constant of the $s\bar{s}\phi$ vertex,
and $F_{\alpha\beta}^{\phi}(x)$ is the $\phi$ field's strength tensor.
Substituting (\ref{eq:polar-vector}) into Eq. (\ref{eq:coalescence-kernel})
and assuming $u^{\mu}=(1,0,0,0)$ in the meson's rest frame, we obtain
the spin alignment of the $\phi$ meson
\begin{align}
 & \rho_{00}^{\phi}(x,{\vec p})\nonumber \\
\approx & \frac{1}{3}-\frac{4g_{\phi}^{2}}{M_{\phi}^{2}T_{\text{h}}}C_{1}\left[\frac{1}{3}{\vec B}_{\phi}^{\prime}\cdot{\vec B}_{\phi}^{\prime}-({\vec\epsilon}_{0}\cdot{\vec B}_{\phi}^{\prime})^{2}\right]\nonumber \\
 & -\frac{4g_{\phi}^{2}}{M_{\phi}^{2}T_{\text{h}}}C_{2}\left[\frac{1}{3}{\vec E}_{\phi}^{\prime}\cdot{\vec E}_{\phi}^{\prime}-({\vec\epsilon}_{0}\cdot{\vec E}_{\phi}^{\prime})^{2}\right],\label{eq:spin-alignment}
\end{align}
where $M_{\phi}$ is the mass of the $\phi$ meson, and ${\vec B}_{\phi}^{\prime}$ and ${\vec E}_{\phi}^{\prime}$ are
electric and magnetic components of the $\phi$ field in the meson's
rest frame respectively. Here ${\vec\epsilon}_{0}$ denotes the spin quantization
direction. The coefficients $C_{1}$ and $C_{2}$ depend only on quark's
and meson's masses,
\begin{eqnarray}
C_{1} & = & \frac{8M_{s}^{4}+16M_{s}^{2}M_{\phi}^{2}+3M_{\phi}^{4}}{120M_{s}^{2}(M_{\phi}^{2}+2M_{s}^{2})},\nonumber \\
C_{2} & = & \frac{8M_{s}^{4}-14M_{s}^{2}M_{\phi}^{2}+3M_{\phi}^{4}}{120M_{s}^{2}(M_{\phi}^{2}+2M_{s}^{2})}.
\end{eqnarray}
We note that the momentum dependence can be obtained by expressing
${\vec B}_{\phi}^{\prime}$ and ${\vec E}_{\phi}^{\prime}$ in Eq. (\ref{eq:spin-alignment})
in terms of laboratory frame fields through Lorentz transformation.
\begin{eqnarray} \label{eq:Lorentz-trans}
{\vec B}_{\phi}^{\prime} & = & \gamma{\vec B}_{\phi}-\gamma{\vec v}\times{\vec E}_{\phi}+(1-\gamma)\frac{{\vec v}\cdot{\vec B}_{\phi}}{v^{2}}{\vec v},\nonumber \\
{\vec E}_{\phi}^{\prime} & = & \gamma{\vec E}_{\phi}+\gamma{\vec v}\times{\vec B}_{\phi}+(1-\gamma)\frac{{\vec v}\cdot{\vec E}_{\phi}}{v^{2}}{\vec v},
\end{eqnarray}
where $\gamma=E_{{\vec p}}^{\phi}/M_{\phi}$ is the Lorentz factor
and ${\vec v}={\vec p}/E_{{\vec p}}^{\phi}$ is the velocity of the
$\phi$ meson. Then $\rho_{00}^{\phi}$ can be rewritten as
\begin{eqnarray}
\rho_{00}^{\phi}(x,{\vec p}) & \approx & \frac{1}{3}-\frac{4g_{\phi}^{2}}{3M_{\phi}^{4}T_{\mathrm{h}}^{2}}\sum_{i=1,2,3}I_{B,i}({\vec p})({\vec B}_{i}^{\phi})^{2}\nonumber \\
 &  & -\frac{4g_{\phi}^{2}}{3M_{\phi}^{4}T_{\mathrm{h}}^{2}}\sum_{i=1,2,3}I_{E,i}({\vec p})({\vec E}_{i}^{\phi})^{2},\label{eq:rho00-eb}
\end{eqnarray}
where $I_{B,i}({\vec p})$ and $I_{E,i}({\vec p})$ are momentum functions.
%\textcolor{red}
{These functions are derived by substituting transformations (\ref{eq:Lorentz-trans}) into Eq. (\ref{eq:spin-alignment}) and then separating different field components. For the global spin alignment, the explicit expression of $I_{B,i}$ reads \cite{Sheng:2022ffb}}
\begin{eqnarray}
I_{B,i}(p)&=&C_1\left[(E_p^\phi)^2+\left(\frac{6E_p^\phi (p_2)^2}{m_\phi+E_p^\phi}-3(E_p^\phi)^2\right)\delta_{2,i}\right.\nonumber\\
&&\hspace{-0.5cm}\left.-\left(1+\frac{3(p_2)^2}{(m_\phi+E_p^\phi)^2}\right)(p_i)^2\right] \nonumber\\
&&\hspace{-0.5cm}+C_2\left[\vec{p}\cdot\vec{p}-(p_i)^2-3\left(\sum_{j=1,2,3}\epsilon_{2ij}p_j\right)^2\right],
\end{eqnarray}
%\textcolor{red}
{where $\epsilon$ in the last line is the rank-3 antisymmetric tensor. The explicit expression of $I_{E,i}(p)$ can be obtained from $I_{B,i}$ by exchanging $C_1$ with $C_2$.} 
One can further take the space and momentum average of $\rho_{00}^{\phi}(x,{\vec p})$
\begin{align}
 & \left\langle \rho_{00}^{\phi}(x,{\vec p})\right\rangle _{x,{\vec p}}\nonumber \\
\approx & \frac{1}{3}-\frac{4g_{\phi}^{2}}{3M_{\phi}^{4}T_{\mathrm{h}}^{2}}\sum_{i=1,2,3}\left\langle I_{B,i}({\vec p})\right\rangle _{{\vec p}}\left\langle ({\vec B}_{i}^{\phi})^{2}\right\rangle _{x}\nonumber \\
 & -\frac{4g_{\phi}^{2}}{3M_{\phi}^{4}T_{\mathrm{h}}^{2}}\sum_{i=1,2,3}\left\langle I_{E,i}({\vec p})\right\rangle _{{\vec p}}\left\langle ({\vec E}_{i}^{\phi})^{2}\right\rangle _{x},\label{eq:rho00-xp}
\end{align}
where the momentum average is defined as
\begin{equation}
\left\langle O({\vec p})\right\rangle_{\vec p}=\frac{\int d^{3}{\vec p}(E_{{\vec p}}^{\phi})^{-1}O({\vec p})f_{\phi}({\vec p})}{\int d^{3}{\vec p}(E_{\vec p}^{\phi})^{-1}f_{\phi}({\vec p})},\label{eq:momentum-average}
\end{equation}
if we want to obtain momentum-integrated data for $\rho_{00}^{\phi}$. 
Here $f_{\phi}({\vec p})$ is its momentum distribution which may
contain information about collective flows such as $v_{1}$ and $v_{2}$. 
If we want to obtain the transverse momentum spectra of $\rho_{00}^{\phi}$,
we have to integrate over the azimuthal angle and rapidity and keep
$p_{T}$, i.e. to replace $d^{3}{\vec p}/E_{p}^{\phi}$ in (\ref{eq:momentum-average})
by $dyd\varphi$. We see in Eq. (\ref{eq:momentum-average}) that the average $\rho_{00}^{\phi}$ 
depends on the space average field squared which quantifies the field fluctuation.       
The theoretical results for $\left\langle \rho_{00}^{\phi}\right\rangle $
as functions of transverse momenta, collision energies and centralities
are presented in Ref. \cite{Sheng:2022wsy}, which are in a good agreement
with recent STAR data \cite{STAR:2022fan}.

\section{Linear response theory for spin alignment of vector mesons in thermal media}

In this section, we will show how to calculate the spin alignment
of the vector meson from the Kubo formula in linear response theory
\cite{zubarev1996statistical,zubarev1997statistical,Zubarev_1979,Kapusta:2006pm}
in thermalized QGP. The detailed discussion of this topic is in Ref.
\cite{Dong:2023cng}.

The Closed-Time-Path (CTP) formalism is a field-theory based method
for many-particle systems in off-equilibrium as well in equilibrium
\cite{Chou:1984es,Blaizot:2001nr,Wang:2001dm,Berges:2004yj,Cassing:2008nn,Crossley:2015evo}.
When it is used for systems in equilibrium, it is actually the real
time formalism of the thermal field theory \cite{Kapusta:2006pm,Kapusta:2023eix}.
Wigner functions can be obtained from two-point Green's functions
on the CTP \cite{Sheng:2021kfc,Sheng:2022ffb,Hidaka:2022dmn,Dong:2023cng,Becattini:2024uha}. 
The on-shell Wigner function for the vector meson is related
to its MVSD (proportional to the spin density matrix) and defined
as
\begin{align}
W^{\mu\nu}(x,p_{\mathrm{on}})= & \frac{E_{p}}{\pi}\int_{0}^{\infty}dp_{0}G_{<}^{\mu\nu}(x,p)\nonumber \\
= & \sum_{m_{1},m_{2}}\epsilon^{\mu}\left(m_{1},{\vec p}\right)\epsilon^{\nu\ast}\left(m_{2},{\vec p}\right)f_{m_{1}m_{2}}^{V}(x,{\vec p}),\label{eq:wigner-decomp}
\end{align}
where $G_{<}^{\mu\nu}(x,p)$ is the ``$+-$'' component of two-point
Green's functions on the CTP, and $f_{V}=\mathrm{Tr}(f_{V})\hat{\rho}^{V}$
with $f_{V}\equiv f_{m_{1}m_{2}}^{V}$ being the short-hand notation
for the MVSD, $\mathrm{Tr}(f_{V})=\sum_{m}f_{mm}^{V}$ being the trace
of the MVSD, and $\hat{\rho}_{V}$ being the spin density matrix in
Eq. (\ref{mvsd-decomp}). One can check that $W^{\mu\nu}(x,p_{\mathrm{on}})$
is always transverse to the on-shell momentum, $p_{\mu}^{\mathrm{on}}W^{\mu\nu}(x,p_{\mathrm{on}})=0$.
The on-shell Wigner function can be decomposed into the scalar ($\mathcal{S}$),
polarization ($W^{[\mu\nu]}$) and tensor polarization ($\mathcal{T}^{\mu\nu}$)
parts as \cite{Sheng:2023chinphyb,Becattini:2024uha}
\begin{align}
W^{\mu\nu}(x,p_{\mathrm{on}})= & W^{[\mu\nu]}+W^{(\mu\nu)}\nonumber \\
= & -\frac{1}{3}\Delta^{\mu\nu}(p_{\mathrm{on}})\mathcal{S}+W^{[\mu\nu]}+\mathcal{T}^{\mu\nu},
\label{eq:wigner-decomp-1}
\end{align}
where each part is defined as
\begin{align}
W^{[\mu\nu]}\equiv & \frac{1}{2}(W^{\mu\nu}-W^{\nu\mu}),\nonumber \\
W^{(\mu\nu)}\equiv & \frac{1}{2}(W^{\mu\nu}+W^{\nu\mu}),\nonumber \\
\mathcal{T}^{\mu\nu}\equiv & W^{(\mu\nu)}+\frac{1}{3}\Delta^{\mu\nu}(p_{\mathrm{on}})\mathcal{S}.
\end{align}
With Eq. (\ref{eq:wigner-decomp-1}) one can show that both $W^{[\mu\nu]}$
and $\mathcal{T}^{\mu\nu}$ are traceless, $g_{\mu\nu}W^{[\mu\nu]}=g_{\mu\nu}\mathcal{T}^{\mu\nu}=0$.
Using Eqs. (\ref{mvsd-decomp}) and (\ref{eq:wigner-decomp}),
we obtain
\begin{align}
\mathcal{S}= & \mathrm{Tr}(f_{V})=-\Delta^{\mu\nu}(p_{\mathrm{on}})W_{\mu\nu},\nonumber \\
W^{[\mu\nu]}= & \frac{1}{2}\mathrm{Tr}(f_{V})\sum_{\lambda_{1},\lambda_{2}}\epsilon^{\mu}\left(\lambda_{1},{\vec p}\right)\epsilon^{\nu\ast}\left(\lambda_{2},{\vec p}\right)P_{i}\Sigma_{\lambda_{1}\lambda_{2}}^{i},\nonumber \\
\mathcal{T}^{\mu\nu}= & \mathrm{Tr}(f_{V})\sum_{\lambda_{1},\lambda_{2}}\epsilon^{\mu}\left(\lambda_{1},{\vec p}\right)\epsilon^{\nu\ast}\left(\lambda_{2},{\vec p}\right)T_{ij}\Sigma_{\lambda_{1}\lambda_{2}}^{ij}.\label{eq:swt-mvsd}
\end{align}
We see that $W^{[\mu\nu]}$ is related to $P_{i}$ while $\mathcal{T}^{\mu\nu}$
is related to $T_{ij}$.

We can extract $f_{00}\propto\rho_{00}$ by projecting
\begin{equation}
L^{\mu\nu}(p_{\mathrm{on}})=\epsilon^{\mu,*}\left(0,{\vec p}\right)\epsilon^{\nu}\left(0,{\vec p}\right)+\frac{1}{3}\Delta^{\mu\nu}(p_{\mathrm{on}}),\label{eq:l-munu-on}
\end{equation}
onto $W^{\mu\nu}$ in Eq. (\ref{eq:wigner-decomp}) as
\begin{align}
 & L_{\mu\nu}(p_{\mathrm{on}})W^{\mu\nu}\nonumber \\
= & \sum_{\lambda_{1},\lambda_{2}}L_{\mu\nu}(p_{\mathrm{on}})\epsilon^{\mu}\left(\lambda_{1},{\vec p}\right)\epsilon^{\nu\ast}\left(\lambda_{2},{\vec p}\right)f_{\lambda_{1}\lambda_{2}}^{V}(x,{\vec p})\nonumber \\
= & f_{00}^{V}(x,{\vec p})+\frac{1}{3}\sum_{\lambda_{1},\lambda_{2}}\epsilon^{\mu}\left(\lambda_{1},{\vec p}\right)\epsilon_{\mu}^{\ast}\left(\lambda_{2},{\vec p}\right)f_{\lambda_{1}\lambda_{2}}^{V}(x,{\vec p})\nonumber \\
= & f_{00}^{V}(x,{\vec p})-\frac{1}{3}\mathrm{Tr}(f_{V}).\label{eq:l-munu}
\end{align}
In (\ref{eq:l-munu-on}), $\epsilon^{\mu}(0,{\vec p})$ is the polarization
vector along the spin quantization direction. With the first line
of Eq. (\ref{eq:swt-mvsd}) and Eq. (\ref{eq:l-munu}), we obtain
\begin{equation}
\frac{L_{\mu\nu}(p_{\mathrm{on}})W^{\mu\nu}}{-\Delta^{\mu\nu}(p_{\mathrm{on}})W_{\mu\nu}}=\frac{f_{00}^{V}(x,{\vec p})}{\mathrm{Tr}[f_{V}(x,{\vec p})]}-\frac{1}{3}=\rho_{00}-\frac{1}{3}.\label{eq:rho-00-1/3}
\end{equation}
The above formula relates the Wigner function to $\rho_{00}$.

The medium effects can be described by retarded and advanced two-point
Green's functions through spectral functions. For vector mesons interacting
with thermal quarks, the spectral function can be defined through
the imaginary part of the retarded two-point Green's function (propagator
in momentum space) as
\begin{equation}
\mathrm{Im}\widetilde{G}_{R}^{\mu\nu}(p)=\Delta_{T}^{\mu\nu}\rho_{T}(p)+\Delta_{L}^{\mu\nu}\rho_{L}(p).\label{eq:im-gr}
\end{equation}
The definition of two-point Green's functions $G$ and $\Sigma$ differs
by a factor $i=\sqrt{-1}$ from the usual one in quantum field theory,
which are related by $G=i\widetilde{G}$ and $\Sigma=i\widetilde{\Sigma}$.
In Eq. (\ref{eq:im-gr}), we defined
\begin{align}
\Delta_{T}^{\mu\nu}= & -g^{\mu0}g^{\nu0}+g^{\mu\nu}+\frac{\mathbf{p}^{\mu}\mathbf{p}^{\nu}}{|{\vec p}|^{2}},\nonumber \\
\Delta_{L}^{\mu\nu}= & \Delta^{\mu\nu}-\Delta_{T}^{\mu\nu}\equiv g^{\mu\nu}-\frac{p^{\mu}p^{\nu}}{p^{2}}-\Delta_{T}^{\mu\nu},\label{eq:projectors}
\end{align}
as the transverse and longitudinal projector respectively with $\mathbf{p}^{\mu}=(0,{\vec p})$,
and $\rho_{T,L}$ are spectral functions in the transverse and longitudinal
directions given by \cite{Dong:2023cng}
\begin{align}
\rho_{T}(p)= & -\mathrm{Im}\frac{1}{p^{2}-m^{2}+\widetilde{\Sigma}_{\perp}(p)+i\,\mathrm{sgn}(p_{0})\varepsilon},\nonumber \\
\rho_{L}(p)= & -\mathrm{Im}\frac{1}{p^{2}-m^{2}+\frac{p^{2}}{|\mathbf{p}|^{2}}\widetilde{\Sigma}_{00}(p)+i\,\mathrm{sgn}(p_{0})\varepsilon},\label{eq:rho-tl}
\end{align}
where $\widetilde{\Sigma}_{\perp}$ and $\widetilde{\Sigma}_{00}$
are from $\widetilde{\Sigma}_{R}^{\mu\nu}$: $\widetilde{\Sigma}_{\perp}\equiv-(1/2)\Delta_{\mu\nu}^{T}\widetilde{\Sigma}_{R}^{\mu\nu}$
and $\widetilde{\Sigma}_{00}=\widetilde{\Sigma}_{R}^{00}$, $\mathrm{sgn}(p_{0})$
is the sign of $p_{0}$, and $\varepsilon$ is an infinitesimal positive
number. One can check in Eq. (\ref{eq:projectors}) that $p_{\mu}\Delta_{T}^{\mu\nu}=p_{\mu}\Delta_{L}^{\mu\nu}=0$.
In Eq. (\ref{eq:rho-tl}), one can verify that the real parts of $\widetilde{\Sigma}_{\perp}$
and $\widetilde{\Sigma}_{00}$ contribute to the mass correction while
the imaginary parts of $\widetilde{\Sigma}_{\perp}$ and $\widetilde{\Sigma}_{00}$
determines the width or life-time of the quasi-particle mode.

The two-point function $G_{<}^{\mu\nu}$ is related to $G_{A}^{\mu\nu}$
and $G_{R}^{\mu\nu}$ through \cite{fetter2003quantum,zubarev1997statistical}
\begin{eqnarray}
G_{<}^{\mu\nu}(p) & = & in_{B}(p_{0})\left[\widetilde{G}_{R}^{\mu\nu}(p)-\widetilde{G}_{A}^{\mu\nu}(p)\right]\nonumber \\
 & = & -2n_{B}(p_{0})\mathrm{Im}\widetilde{G}_{R}^{\mu\nu}(p),\label{eq:less_R}
\end{eqnarray}
where $n_{B}(p_{0})=1/(e^{\beta p_{0}-\beta\mu_{V}}-1)$ is the Bose-Einstein
distribution with the inverse temperature $\beta\equiv1/T$ and the
vector meson's chemical potential $\mu_{V}$. Inserting Eq. (\ref{eq:less_R})
into (\ref{eq:wigner-decomp}) we obtain the on-shell Wigner function
$W^{\mu\nu}(p_{\mathrm{on}})$ from which the spin alignment $\rho_{00}$
can be extracted through Eq. (\ref{eq:rho-00-1/3}).

For free vector mesons, the spectral functions are $\rho_{T}^{(0)}=\rho_{L}^{(0)}=\pi\mathrm{sgn}(p_{0})\delta(p^{2}-m^{2})$,
which give $G_{<}^{\mu\nu}(p)$ and $\mathrm{Im}\widetilde{G}_{R}^{\mu\nu}(p)$
for the free vector meson following Eqs. (\ref{eq:im-gr}) and (\ref{eq:less_R}).

The non-equilibrium correction to $G_{<}^{\mu\nu}(p)$ can be calculated
through the Kubo formula in linear response theory. The Kubo formula
has been derived in Zubarev's approach to non-equilibrium density
operator \cite{Zubarev_1979,Hosoya:1983id,Becattini:2019dxo}. Detailed
derivation is given in Ref. \cite{Becattini:2019dxo}. Considering
the thermal shear tensor $\xi_{\mu\nu}=1/2(\partial_{\mu}\beta_{\nu}+\partial_{\nu}\beta_{\mu})$
as a perturbation from local equilibrium, the next-to-leading order
correction of $G_{<}^{\mu\nu}$ can be written as \cite{Dong:2023cng}
\begin{align}
 & \delta G_{<}^{\mu\nu}(x,p)\nonumber \\
= & 4T\lim_{K^{\mu}\rightarrow0}\frac{\partial}{\partial K_{0}}\mathrm{Im}\int\frac{dp_{1}^{0}dp_{2}^{0}}{2\pi}\nonumber \\
 & \times\frac{n_{B}(p_{1}^{0})-n_{B}(p_{2}^{0})}{p_{1}^{0}-p_{2}^{0}+K^{0}+i\epsilon}\delta\left(p^{0}-\frac{p_{1}^{0}+p_{2}^{0}}{2}\right)\xi_{\gamma\lambda}\nonumber \\
 & \times\sum_{a,b=L,T}\rho_{a}(p_{1})\rho_{b}(p_{2})I_{ab}^{\mu\nu\gamma\lambda}(p_{1},p_{2}),\label{eq:NLO}
\end{align}
where $p_{1}=(p_{1}^{0},{\vec p}-{\vec K}/2),p_{2}=(p_{2}^{0},{\vec p}+{\vec K}/2)$.
The tensor $I_{ab}^{\mu\nu\gamma\lambda}(p_{1},p_{2})$ is expressed as \cite{Dong:2023cng}
\begin{align}
 & I_{ab}^{\mu\nu\gamma\lambda}(p_{1},p_{2})\nonumber \\
= & (p_{1}^{\lambda}p_{2}^{\gamma}+p_{1}^{\gamma}p_{2}^{\lambda})\Delta_{a,\alpha}^{\nu}(p_{1})\Delta_{b}^{\mu\alpha}(p_{2})\nonumber \\
 & +(p_{1,\alpha}p_{2}^{\alpha}-m_{V}^{2})\nonumber \\
 & \times\left[\Delta_{a}^{\gamma\nu}(p_{1})\Delta_{b}^{\mu\lambda}(p_{2})+\Delta_{a}^{\lambda\nu}(p_{1})\Delta_{b}^{\mu\gamma}(p_{2})\right]\nonumber \\
 & -\left[p_{1}^{\gamma}p_{2}^{\alpha}\Delta_{a,\alpha}^{\nu}(p_{1})\Delta_{b}^{\mu\lambda}(p_{2})\right.\nonumber \\
 & \left.+p_{2}^{\gamma}p_{1}^{\alpha}\Delta_{a}^{\lambda\nu}(p_{1})\Delta_{b,\alpha}^{\mu}(p_{2})\right]\nonumber \\
 & -\left[p_{1,\alpha}p_{2}^{\lambda}\Delta_{a}^{\gamma\nu}(p_{1})\Delta_{b}^{\mu\alpha}(p_{2})\right.\nonumber \\
 & \left.+p_{1}^{\lambda}p_{2,\alpha}\Delta_{a}^{\alpha\nu}(p_{1})\Delta_{b}^{\mu\gamma}(p_{2})\right]\nonumber \\
 & -g^{\gamma\lambda}\left[g_{\beta\alpha}(p_{2,\rho}p_{1}^{\rho}-m_{V}^{2})-p_{1,\beta}p_{2,\alpha}\right]\nonumber \\
 & \times\Delta_{a}^{\alpha\nu}(p_{1})\Delta_{b}^{\mu\beta}(p_{2}),\label{eq:I_ab}
\end{align}
where $\Delta_{T}^{\mu\nu}(p)$ and $\Delta_{L}^{\mu\nu}(p)$ are
given by Eq. (\ref{eq:projectors}). Then the spin alignment of vector
mesons can be obtained from Eq. (\ref{eq:rho-00-1/3}) through Eqs.
(\ref{eq:wigner-decomp}) and (\ref{eq:NLO}). 

\begin{figure}
\includegraphics[scale=0.11]{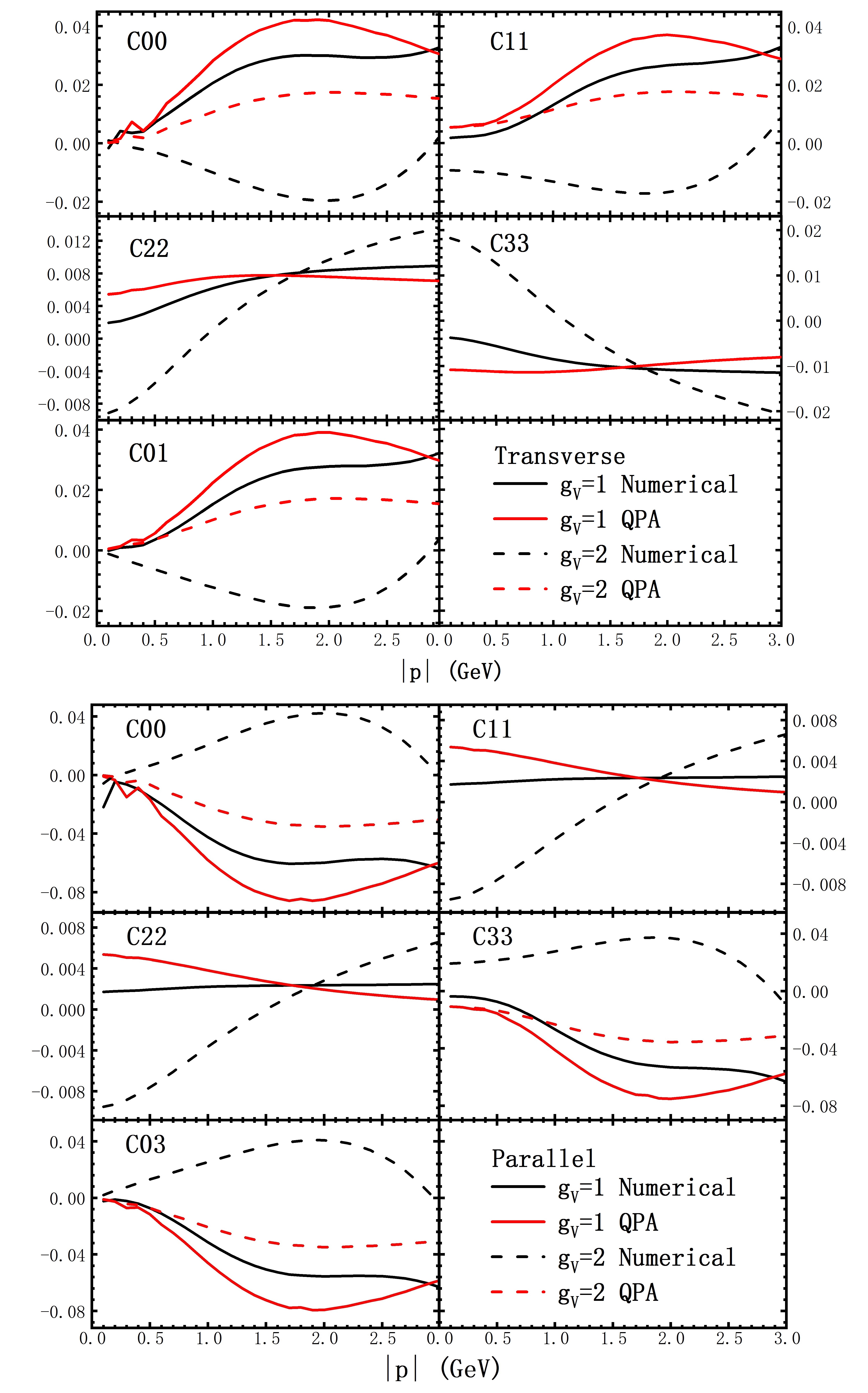}\\
\caption{The full numerical and QPA results for $C^{\mu\nu}$ as functions
of vector mesons' momenta. In the upper panel the momentum is perpendicular
to the spin quantization direction $z$, while in the lower panel the momentum
is parallel to the spin quantization direction. \label{fig:The-numerical-results}}
\end{figure}

Under the quasi-particle approximation (QPA), the self-energies and
spectral functions can be calculated analytically by expanding $p_{0}$
in powers of $\delta p_{0}=p_{0}-E_{p}^{V}$. This expansion requires
$\Delta E/E_{p}^{V}\sim\Gamma/E_{p}^{V}\sim\epsilon\ll1$, where $\Delta E$
and $\Gamma$ are the mass shift and width of the vector meson respectively.
Then the spin alignment can be expressed as 
\begin{equation}
\delta\rho_{00}({\vec p})=\delta\rho_{00}^{(\xi=0)}({\vec p})+\xi_{\mu\nu}C^{\mu\nu}({\vec p}),
\end{equation}
where $\delta\rho_{00}^{(\xi=0)}$ is the spin alignment independent of 
the shear stress tensor and $C^{\mu\nu}$ is the dimensionless tensor coefficient 
in the shear stress term. The results under the QPA are compared with full numerical results
are shown in Fig. (\ref{fig:The-numerical-results}). 
For the values of parameters we choose, the magnitudes of 
$\delta\rho_{00}^{(\xi=0)}$ and $C^{\mu\nu}({\vec p})$ are $\mathcal{O}(10^{-3})$ 
and $\mathcal{O}(10^{-2}\sim 10^{-3})$, respectively.  
As a consequence, if the thermal shear tensor is $\mathcal{O}(10^{-2})$, 
the contribution from the thermal shear tensor to the $\phi$ meson's spin alignment 
is $\mathcal{O}(10^{-4}\sim10^{-5})$, which is much smaller 
than the contribution from the strong force fields \cite{Sheng:2022wsy}.

\section{Vector meson's spin alignments in quark fragmentation} \label{sec:FFs}

In high energy reactions, hadron production is normally described by fragmentation functions (FFs). 
The hadron‘s Polarization can be realized in terms of polarized FFs~\cite{Chen:2016moq,Chen:2016iey,Chen:2020pty}.
These FFs are defined via quark-quark correlators. 
Among different high energy reactions, the clean and ideal place to study FFs is in high energy $e^+e^-$ collisions. 
In Ref.~\cite{Chen:2016moq}, the complete decomposition of quark-quark correlators for spin-1 hadrons with systematic results for hadron polarization in terms of FFs up to twist-3 in $e^+e^-\to V\pi X$ have been presented. 
Experimental measurements have been carried out many years ago for the longitudinal polarization of $\Lambda$ hyperons~\cite{ALEPH:1996oew,OPAL:1997oem}
and for the spin alignments of vector mesons~\cite{DELPHI:1997ruo,OPAL:1997vmw,OPAL:1997nwj,OPAL:1999hxs} in the inclusive production process $e^+e^-\to hX$ at LEP. Significant effects have been observed.
These experimental data have attracted many phenomenological studies~\cite{Anselmino:1997ui,Anselmino:1998jv,Anselmino:1999cg,Xu:2001hz,Xu:2003fq,Chen:2016moq,Chen:2016iey,Chen:2020pty}
and parameterization of corresponding FFs has been obtained~\cite{Chen:2016moq,Chen:2016iey,Chen:2020pty}.
In this section we will summarize these results.

\subsection{FFs from quark-quark correlator}
For the fragmentation of the quark (or anti-quark), the quark-quark correlator 
as a $4 \times 4$ matrix in Dirac indices is defined by 
\begin{align}
\hat\Xi(k;p,S)= & \sum_X \int  \frac{ d^4\xi }{2\pi} e^{-ik \xi}  \langle hX|\bar\psi(\xi) \mathcal{L}(\xi;\infty) |0\rangle \nonumber\\
& \times \langle 0| \mathcal{L}^\dag (0;\infty) \psi(0) |hX\rangle, \label{qq-correlator}
\end{align}
where $k$ and $p$ are four-momenta of the quark and hadron respectively, 
$S$ denotes the hadron's spin, and $\mathcal{L}(\xi;\infty)$ is the gauge link  defined as 
\begin{equation}
\mathcal{L}(\xi;\infty) =P\mathrm{exp}
\left[ig\int_{\xi^-}^\infty d\eta^-A^+(\eta^-;\xi^+,\vec{\xi}_\perp)\right],
\end{equation}
which guarantees the gauge invariance of FFs. 
In the rest part of the paper, for notational simplicity, we suppress the gauge link in forthcoming equations. 
In this case, the field operator $\psi(\xi)$ just stands for $\mathcal{L}^\dag (\xi;\infty) \psi(\xi)$. 
%One can obtain $\hat\Xi(z;p,S)$ by integrating over $k^-$ and $\vec{k}_\perp$ from $\hat\Xi(k;p,S)$. 

% corrected by QW, 8:00, 2024.5.4

The FFs are obtained from $\hat\Xi(k;p,S)$ in the following way.
First, we expand $\hat\Xi(k;p,S)$ in terms of 
$\Gamma$ matrices as 
%=(1,\gamma_5, \gamma^\alpha, \gamma_5\gamma^\alpha, \sigma^{\alpha\beta})$  
\begin{align}
\hat\Xi(k;&p,S) = \frac{1}{2} \Bigl[ \Xi(k;p,S) + i\gamma_5 \tilde\Xi(k;p,S) \nonumber\\
& + \gamma^\alpha \Xi_\alpha(k;p,S) + \gamma_5\gamma^\alpha \tilde\Xi_\alpha(k;p,S) \nonumber\\
& + i\sigma^{\alpha\beta}\gamma_5 \Xi_{\alpha\beta}(k;p,S) \Bigr], \label{XiExpansion}
\end{align}
where coefficients are all real functions and 
are Lorentz scalar, pseudo-scalar, vector, axial-vector and tensor respectively.
Then we expand these coefficient functions according to their respective Lorentz transformation properties
in terms of basic Lorentz covariants multiplied by scalar functions which are constructed from available variables. These scalar functions are the FFs. 

% corrected by QW, 8:40, 2024.5.4

Clearly, basic Lorentz covariants to be constructed depend strongly on what variables that we have at hand.
Besides $p$ and $k$, we have variables for spin states which are different for spin-0, 1/2 or 1 hadrons.
%We therefore obtain different results for hadrons with different spins.
For this purpose, we need to decompose the spin density matrix in terms of known operators multiplied by Lorentz covariants. 
For spin-1/2 hadrons, the spin density matrix $\rho$ is decomposed as in Eq.~(\ref{eq:rho1/2}),
but $\vec{S}$ in the rest frame of the hadron should be generalized to a covariant form $S=(0,\vec{S})$.
For spin-1 hadrons, the $3\times 3$ density matrix $\hat\rho$ is usually decomposed as in Eq. (\ref{mvsd-decomp}). 
The tensor polarization $T^{ij}={\rm Tr}(\rho \Sigma^{ij})$ can be parameterized as 
\begin{align}
%\mathbf{T}
[T]=\frac{1}{2}
\left(
\begin{array}{ccc}
-\frac{2}{3}S_{LL} + S_{TT}^{xx} & S_{TT}^{xy} & S_{LT}^x  \\
S_{TT}^{xy}  & -\frac{2}{3} S_{LL} - S_{TT}^{xx} & S_{LT}^{y} \\
S_{LT}^x & S_{LT}^{y} & \frac{4}{3} S_{LL}
\end{array}
\right).
\label{spintensor}
\end{align}
We see that, for spin-1 hadrons, besides the polarization vector $S$, there is also a tensor polarization part.
The polarization vector is similar to that for spin-1/2 hadrons.
The tensor polarization part has five components, namely 
a Lorentz scalar $S_{LL}$, a Lorentz vector $S_{LT}^\mu = (0, S_{LT}^x, S_{LT}^y,0)$
and a Lorentz tensor $S_{TT}^{\mu\nu}$ that has two components
$S_{TT}^{xx} = -S_{TT}^{yy}$ and $S_{TT}^{xy} = S_{TT}^{yx}$.
The vector meson's spin alignment $\rho_{00}$ is only related to $S_{LL}$ through $\rho_{00}=(1-2S_{LL})/3$. 

% corrected by QW, 9:40, 2024.5.4

The quark-quark correlator given by Eq.~(\ref{qq-correlator}) is un-integrated, i.e. it depends on the four-momentum $k$.
If we consider three- or one-dimensional FFs, we need to integrate it over $k^-$ and $\vec k_\perp$.
We take one-dimensional FFs as an example. 
In this case, after integrating over $k^-$ and $\vec k_\perp$, we obtain,
\begin{align}
\hat\Xi(z;p,S) = & \sum_X \int  \frac{ d\xi^- }{2\pi} e^{-ik^+\xi^-} \nonumber\\
& \times  \langle hX|\bar\psi(\xi^-)  |0\rangle \langle 0|  \psi(0) |hX\rangle, \label{eq:1qq-correlator}
\end{align}
where $z\equiv p^+/k^+$. After Lorentz decomposition, we obtain terms related to $S_{LL}$ as,
\begin{align}
z\Xi_\alpha(z;p,S) = &p^+ \bar n_\alpha [D_1(z) + S_{LL} D_{1LL}(z)] + \nonumber\\
&+ {\rm power~suppressed~terms}. \label{eq:D1LL}
\end{align}
We can obtain the expression for $D_1(z) + S_{LL} D_{1LL}(z)$ by reversely solving Eqs.~(\ref{eq:1qq-correlator}) and (\ref{eq:D1LL}), 
\begin{align}
D_1(z)+ & S_{LL} D_{1LL}(z)=\sum_X \int  \frac{ zd\xi^- }{2\pi p^+} e^{-ik^+\xi^-}  \nonumber\\
& \times \langle hX|\bar\psi(\xi^-) \gamma^+ |0\rangle \langle 0|  \psi(0) |hX\rangle, 
\label{eq:expD1LL}
\end{align}
For comparison, we present the corresponding equations for $G_{1L}$ that describes the longitudinal spin transfer in the fragmentation process
\begin{align}
z\tilde\Xi_\alpha(z;p,S) = & \lambda p^+ \bar n_\alpha G_{1L} (z) +{\rm power~suppressed~terms} \nonumber\\
\lambda G_{1L}(z)=& \sum_X \int  \frac{ zd\xi^- }{2\pi p^+} e^{-ik^+\xi^-}  \nonumber\\
& \times \langle hX|\bar\psi(\xi^-) \gamma^+\gamma_5 |0\rangle \langle 0|  \psi(0) |hX\rangle. 
\label{eq:expG1L}
\end{align}
The difference between $D_1(z) + S_{LL} D_{1LL}(z)$ in Eq. (\ref{eq:expD1LL}) and $\lambda G_{1L}$ in Eq. (\ref{eq:expG1L}) is that $\gamma^+$ is involved in the former while $\gamma^+\gamma_5$ is 
involved in the latter.  

% corrected by QW, 10:00, 2024.5.4

\subsection{Vector meson's spin alignments from FFs}

From Eqs.~(\ref{eq:expD1LL}), we see clearly that, similar to the well-known unpolarized FF $D_1(z)$, $D_{1LL}$ is independent of the quark's polarization because it is a sum over  the quark's spin states 
\begin{align}
&D_1(z) + S_{LL} D_{1LL}(z)=\sum_X \int  \frac{ zd\xi^- }{2\pi p^+} e^{-ik^+\xi^-}  \nonumber\\
& \times \sum_{\lambda_q=L,R} \langle hX|\bar\psi_{\lambda_q} (\xi^-) \gamma^+ |0\rangle \langle 0|  \psi_{\lambda_q}(0) |hX\rangle, \label{eq:spinD1LL}
\end{align}
where $\lambda_q=L,R$ denotes the quark's helicity (or chirality), and $\psi_{L/R}=(1\pm\gamma_5)\psi/2$. In contrast, the result for $G_{1L}$ gives 
\begin{align}
\lambda G_{1L}(z)=&\sum_X \int  \frac{ zd\xi^- }{2\pi p^+} e^{-ik^+\xi^-}  \nonumber\\
\times &[ \langle hX|\bar\psi_L (\xi^-) \gamma^+ |0\rangle \langle 0|  \psi_L(0) |hX\rangle \nonumber\\
&-\langle hX|\bar\psi_R (\xi^-) \gamma^+ |0\rangle \langle 0|  \psi_R(0) |hX\rangle ], \label{eq:spinG1L}
\end{align}
which depends on the quark's spin explicitly.

% corrected by QW, 12:00, 2024.5.4

We can draw an important conclusion from Eq.~(\ref{eq:spinD1LL}) for the spin alignment $\rho_{00}=(1-2S_{LL})/3$ for the vector meson produced in the fragmentation process $q\rightarrow VX$: 
it is determined by $D_{1LL}$ and independent of the initial polarization of the quark.
The conclusion is rather unexpected because the vector meson's spin alignment in high energy reactions was first observed in $e^+e^-\rightarrow VX$ at LEP~\cite{DELPHI:1997ruo,OPAL:1997vmw,OPAL:1997nwj,OPAL:1999hxs}
where initial quarks and anti-quarks are longitudinally polarized.
However, this is consistent with space reflection in fragmentation processes where $\rho_{00}$ is space reflection invariant while the helicity of the initial quark changes the sign. 
This conclusion is rather solid since it follows from the general principle of QCD.
It can also be tested easily in experiments. 
In the following we present numerical results for $e^+e^-\rightarrow VX$ and $pp\rightarrow VX$ as examples. 

% corrected by QW, 12:00, 2024.5.4
\subsection{Vector meson's spin alignments in $e^+e^-\rightarrow VX$}

Suppose that the quark fragmentation mechanism dominates hadron production in $e^+e^-$ annihilation at high energies, we obtain the vector meson's alignment in $e^+e^-\rightarrow VX$ as 
\begin{align}
&\langle S_{LL}\rangle (z,y) = \frac{\sum_q W_q(y)D_{1LL}(z)}{2\sum_q W_q(y)D_1(z)}, \label{eq:eeSll} \\
&\langle \lambda\rangle (z,y)
=  \frac{\sum_q P_q(y) W_q(y) G_{1L}(z)}{\sum_q W_q(y) D_1(z)}, \label{eq:eeLamPol}
\end{align}
where we also show the result of hyperon polarization for comparison. 
In Eqs. (\ref{eq:eeSll}) and (\ref{eq:eeLamPol}), $P_q(y)$ and $W_q(y)$ are the quark polarization and production weight at the vertex of $e^+e^-$ annihilation respectively which are determined by the quark's electric charge, $c_V^e$ and $c_A^e$ in the weak interaction current, and so on (see e.g. ~\cite{Chen:2016iey} for the detailed expressions), $y=l_2 \cdot p_q/[(l_1+l_2)\cdot p_q]$, where $l_1$ and $l_2$ are the four-momenta of the incident $e^-$ and $e^+$ respectively, and $p_q$ is the four-momentum of the produced quark. 

% corrected by QW, 15:00, 2024.5.4

\begin{figure}[!ht]
\includegraphics[width=0.35\textwidth]{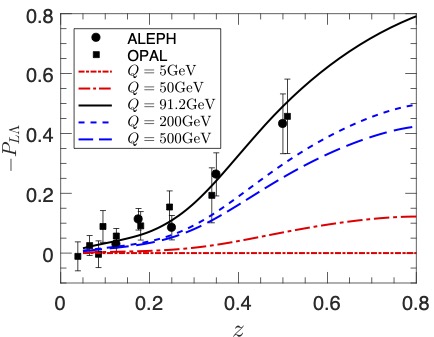}
\caption{Longitudinal polarization of $\Lambda$ in $e^+e^-\to\Lambda X$ at high energies.
The LEP data are taken from Refs.~\cite{ALEPH:1996oew,OPAL:1997oem}.
The solid line is the fit obtained in Ref.~\cite{Chen:2016iey} at the LEP energy while those at other energies are calculated results using the DGLAP equation for FFs and energy dependence of $P_q$. The figure is taken from Ref.~\cite{Chen:2016iey}. }
\label{fig:fit_Lambda}
\end{figure}

\begin{figure}[h!]
\includegraphics[width=0.4\textwidth]{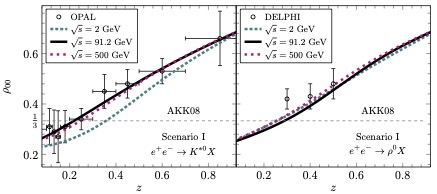}
\caption{The spin alignments of $K^{*0}$ and $\rho^0$ in $e^+e^-\to VX$ at the $Z$-pole
fitted in Ref.~\cite{Chen:2020pty} compared with experimental data~\cite{DELPHI:1997ruo,OPAL:1997vmw}.
The solid line is the fit in Ref.~\cite{Chen:2020pty} at the LEP energy while those at other energies are the results using the DGLAP equation for FFs.
The figure is taken from Ref.~\cite{Chen:2016iey}. } \label{fig:fit_SA}
\end{figure}

The FFs are extracted from data available at a given scale and their scale dependence is determined
by the QCD evolution equation, the DGLAP equation~\cite{Dokshitzer:1977sg,Gribov:1972ri,Altarelli:1977zs,Owens:1978qz,Georgi:1977mg,Uematsu:1978yw,Ravindran:1996ri,Ravindran:1996jd}.
In Ref.~\cite{Chen:2016iey,Chen:2020pty}, such parameterizations of $D_{1LL}$ for vector mesons have been obtained for the first time by fitting available data~\cite{DELPHI:1997ruo,OPAL:1997vmw,OPAL:1997nwj}. 
As comparison, parameterizations of $G_{1L}$ for $\Lambda$ are also given therein~\cite{Chen:2016iey,Chen:2020pty} by fitting
the LEP data~\cite{ALEPH:1996oew,OPAL:1997oem}.
Here, we show the fitted results obtained there~\cite{Chen:2016iey,Chen:2020pty} in Figs.~\ref{fig:fit_Lambda} and \ref{fig:fit_SA}. 

% corrected by QW, 16:30, 2024.5.4

% corrected by QW, 16:30, 2024.5.4

% corrected by QW, 16:30, 2024.5.4

From Figs.~\ref{fig:fit_Lambda} and \ref{fig:fit_SA}, we see a distinct feature that there is a strong energy dependence for $P_{L\Lambda}$, while the energy dependence for $\rho_{00}^{K^*}$ is weak. 
The former comes mainly from the energy dependence of $P_q$ while the latter comes mainly from the QCD evolution of FFs.
%In contrast to $P_{L\Lambda}$, $\rho_{00}$ changes with $Q$ quite weakly and remains sizable even at lower energies.
The relatively rapid change in the energy region around the mass of $Z$ boson comes from the influence of $W_q$. 
This feature can be tested in future experiments.

\subsection{Vector meson's spin alignments in $pp\rightarrow VX$}

\begin{figure}[h!]
\includegraphics[width=0.375\textwidth]{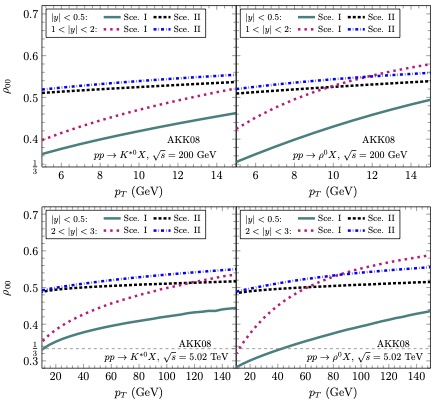}
\caption{Spin alignments of vector mesons in $pp$ collisions at $\sqrt{s}=200$ GeV and $\sqrt{s}=5.02$ TeV
for $K^{*0}$ and $\rho^0$ in two rapidity regions as functions of $p_T$.
This figure is taken from Ref.~\cite{Chen:2020pty}.}
\label{fig:ppSA}
\end{figure}

Assuming the universality of FFs, one can calculate vector meson's spin alignments in other high energy reactions
where the fragmentation mechanism dominates. 
This applies to hadron production at high transverse momentum in $pp$ collisions and deeply inelastic scatterings. In Ref.~\cite{Chen:2020pty}, such calculations have been carried out in $pp$ collisions.
As examples, we show in Fig.~\ref{fig:ppSA} the results obtained there for $K^{*0}$ and $\rho^0$ in two rapidity regions as functions of $p_T$.

We see that $\rho^V_{00}$ for vector mesons deviate from $1/3$ significantly showing large spin alignments. 
Such results can be tested in experiments at RHIC and LHC.

\section{Summary and outlook}

Spin alignments for vector mesons have been observed in high energy reactions~\cite{STAR:2022fan,DELPHI:1997ruo,OPAL:1997vmw,OPAL:1997nwj,OPAL:1999hxs,NOMAD:2006kuc}.
There are two types of spin alignments: global spin alignments in heavy-ion collisions~\cite{STAR:2022fan}
and helicity-basis spin alignments in $e^+e^-$ collisions~\cite{DELPHI:1997ruo,OPAL:1997vmw,OPAL:1997nwj,OPAL:1999hxs}. These measurements show different features of spin alignments arising from different hadronization mechanisms in collisions of heavy-ion and $e^+e^-$.

In high energy heavy-ion collisions in the region of central rapidity and small transverse momentum, 
hadron production is mainly through quark combination or coalescence. 
The global spin alignment in this case not only depends on the polarization of quarks and anti-quarks
but is also sensitive to the local correlation between the polarization of quarks and that of anti-quarks.
Measurements of the global spin alignment in this case provide a good opportunity 
to study the polarization correlation in the sQGP produced in heavy-ion collisions. 
The STAR's measurements of $\phi$ meson's spin alignments~\cite{STAR:2022fan} is consistent with the theoretical description based on $\phi$ vector fields that lead to a strong local correlation between $P_s$ and $P_{\bar s}$~\cite{Yang:2017sdk,Sheng:2020ghv,Sheng:2022ffb,Sheng:2022wsy}. 
The data were used to extract the strength of the local fluctuation in $\phi$ vector fields~\cite{Sheng:2022ffb,Sheng:2022wsy}.

In $e^+e^-$ and $pp$ collisions at high energies, hadron production is dominated by fragmentation with fragmentation functions. 
It has been shown that the spin alignment of vector mesons in the helicity basis is independent of the polarization of the initial quark in the quark fragmentation process. 
With the parameterization of corresponding fragmentation functions the prediction on spin alignments of vector mesons 
in high energy $e^+e^-$ and $pp$ collisions has been made~\cite{Chen:2016moq,Chen:2016iey,Chen:2020pty}.

It is impressive and interesting to see the distinction between the global spin alignments in high energy heavy-ion collisions and in $e^+e^-$ or $pp$ collisions which are dominated by different hadronization mechanisms.  
It is important to extend the measurements to different high energy reactions at different energies to test the universality of these properties. 
Further measurements on global spin alignments of different vector mesons in relativistic heavy ion collisions are expected to be carried out in the near future by the STAR Collaboration at RHIC and ALICE Collaboration at LHC. 
More studies are also planned at GSI, HIAF and NICA in the lower energy regions. 
Also studies in $pp$ collisions are expected by STAR at RHIC, and in $e^+e^-$ by BESIII at BEPC and Belle II at KEK. 
They can also be measured in lepton-nucleon scatterings at future EIC. 
Obviously such studies provide new insights into properties of sQGP and hadronization mechanisms.\\

\section*{Acknowledgements}
This work was supported in part by the National Key Research and Development Program of China under Contract No. 2022YFA1604900,
by the National Natural Science Foundation of China (NSFC) under Contract Nos. 12025501, 11890710, 11890713, 11890714, 12147101 and 12135011,
by the Strategic Priority Research Program of the Chinese Academy of Sciences (CAS) under Grant No. XDB34030102 and by the Natural Science Foundation of Shandong Province.

\bibliographystyle{h-physrev}
\bibliography{ref}

\end{document}